\documentclass[aps,reprint,balancelastpage, superscriptaddress,amsmath,amssymb,floatfix,longbibliography]{revtex4-2}
\usepackage[utf8]{inputenc}

\usepackage{mathtools,esdiff}
\usepackage{braket}
\usepackage{siunitx}
\usepackage{hyperref}
\usepackage[export]{adjustbox}
\usepackage{xcolor}

\newcommand{\im}{\mathrm{i}}
\newcommand{\e}[1]{\mathrm{e}^{#1}}
\newcommand{\T}{\mathrm{T}}
\newcommand{\dif}{\mathrm{d}}
\newcommand{\inv}{^{-1}}
\newcommand{\paren}[1]{\left(#1\right)}
\newcommand{\vbrak}[1]{\left\langle#1\right\rangle}
\newcommand{\sbrak}[1]{\!\left[#1\right]}
\newcommand{\Paren}[1]{\bigl(#1\bigr)}
\newcommand{\Vbrak}[1]{\bigl\langle#1\bigr\rangle}
\newcommand{\Sbrak}[1]{\bigl[#1\bigr]}
\newcommand{\mean}[1]{\vbrak{#1}}
\newcommand{\covar}[2]{\vbrak{\delta#1\,\delta#2}}
\newcommand{\hrho}[1]{\hat\rho\Paren{#1}}
\newcommand{\pump}{_{b}}

\DeclareMathOperator{\sgn}{sgn}
\DeclareMathOperator{\tr}{tr}
\DeclareMathOperator{\diag}{diag}
\DeclareMathOperator{\beamsplitter}{\mathcal{B}}

\DeclareMathOperator{\displace}{\mathcal{V}}
\DeclareMathOperator{\homodyne}{\mathcal{M}}

\begin{document}

\title{Efficient sampling of ground and low-energy Ising spin configurations with a coherent Ising machine}

\author{Edwin~Ng}
\thanks{These authors contributed equally to this work. Email to \{edwin.ng, tatsuhiro.onodera\}@ntt-research.com.}
\affiliation{Physics \& Informatics Laboratories, NTT Research, Inc., Sunnyvale, California 94085, USA}
\affiliation{E.\,L.\,Ginzton Laboratory, Stanford University, Stanford, California 94305, USA}

\author{Tatsuhiro~Onodera}
\thanks{These authors contributed equally to this work. Email to \{edwin.ng, tatsuhiro.onodera\}@ntt-research.com.}
\affiliation{Physics \& Informatics Laboratories, NTT Research, Inc., Sunnyvale, California 94085, USA}
\affiliation{School of Applied and Engineering Physics, Cornell University, Ithaca, New York 14853, USA}

\author{Satoshi~Kako}
\affiliation{Physics \& Informatics Laboratories, NTT Research, Inc., Sunnyvale, California 94085, USA}

\author{Peter~L.~McMahon}
\affiliation{School of Applied and Engineering Physics, Cornell University, Ithaca, New York 14853, USA}

\author{Hideo~Mabuchi}
\affiliation{E.\,L.\,Ginzton Laboratory, Stanford University, Stanford, California 94305, USA}

\author{Yoshihisa~Yamamoto}
\affiliation{Physics \& Informatics Laboratories, NTT Research, Inc., Sunnyvale, California 94085, USA}

\date{\today}

\begin{abstract}
We show that the nonlinear stochastic dynamics of a measurement-feedback-based coherent Ising machine (MFB-CIM) in the presence of quantum noise can be exploited to sample degenerate ground and low-energy spin configurations of the Ising model. We formulate a general discrete-time Gaussian-state model of the MFB-CIM which faithfully captures the nonlinear dynamics present at and above system threshold. This model overcomes the limitations of both mean-field models, which neglect quantum noise, and continuous-time models, which assume long photon lifetimes. Numerical simulations of our model show that when the MFB-CIM is operated in a quantum-noise-dominated regime with short photon lifetimes (i.e., low cavity finesse), homodyne monitoring of the system can efficiently produce samples of low-energy Ising spin configurations, requiring many fewer roundtrips to sample than suggested by established high-finesse, continuous-time models. We find that sampling performance is robust to, or even improved by, turning off or altogether reversing the sign of the parametric drive, but performance is critically reduced in the absence of optical nonlinearity. For the class of MAX-CUT problems with binary-signed edge weights, the number of roundtrips sufficient to fully sample all spin configurations up to the first-excited Ising energy, including all degeneracies, scales with the problem size $N$ as $1.08^N$. At $N=100$ with a few dozen (median $\sim\num{20}$) such desired configurations per instance, we have found median sufficient sampling times of \num{6e6} roundtrips; in an experimental implementation of an MFB-CIM with a \SI{10}{GHz} repetition rate, this corresponds to a wall-clock sampling time of \SI{60}{\milli\second}.
\end{abstract}
\maketitle

\section{Introduction}
For decades, the Ising model has served as a key conceptual bridge between the fields of physics and computation. A host of important combinatorial optimization problems have efficient mappings to the problem of finding ground states of the Ising model~\cite{Lucas2014}, while the simple and highly generic form of the model means that Ising-like interactions are ubiquitous across a diverse array of systems~\cite{Brush1967}. Formally, the Ising model consists of a set of spins $\sigma_i = \pm 1$ with configuration energy given by the Ising Hamiltonian $-\sum_{i\neq j} J_{ij} \sigma_i\sigma_j$, and, in general, the Ising problem of finding spin configurations that minimize this energy is presently intractable on conventional computers~\cite{Barahona1982}. As a result, significant interest has developed towards leveraging physical Ising-like systems as special-purpose computational hardware for tackling problems such as combinatorial optimization, with ongoing research on platforms ranging from quantum annealers built from microwave superconducting circuits~\cite{Johnson2011, Boixo2014} to coherent Ising machines~\cite{Marandi2014b, McMahon2016, Inagaki2016, Inagaki2016b} based on networks of nonlinear optical oscillators among many others~\cite{Mahboob2016, Tianshi2019, Pierangeli2019, Okawachi2020, Chou2019, Cai2020}.

But while combinatorial optimization is often focused on finding just one of the ground-state Ising spin configurations, it is desirable in many applications to obtain many or all degenerate ground-state configurations, and, in some cases, to sample many low-energy configurations as well~\cite{Zhu2019}. Such sampling capability is particularly useful for applications that involve obtaining distributional information about spin configurations in an Ising model, such as estimating the ground-state entropy of a physical simulation with Ising-like interactions or implementing Boltzmann machines as generative models for machine learning~\cite{Hinton2002, Salakhutdinov2007, PerdomoOrtiz2018}. In industrial settings, accessing a pool of candidate solutions to an optimization problem can make processes more efficient and flexible; for example in drug discovery~\cite{Bohacek1996, Lounnas2013, Ogata2010, Sakaguchi2016}, structure-based lead optimization could generate a number of candidate molecules for simultaneous testing. Recently, it has also been pointed out~\cite{Zhengbing2014, Zhengbing2016} that when decomposing large optimization problems into subproblems to be solved separately (e.g., to accommodate hardware limitations), better solutions to the original problem can be constructed using multiple low-energy samples rather than just the optimum for each subproblem. However, an Ising solver designed for combinatorial optimization is not necessarily well suited to sample all ground states and/or low-energy states. For instance, although the commercial quantum annealers by D-Wave Systems have shown success in finding ground states of the Ising problem, their principle of operation can lead to an exponential bias in the distributions of degenerate ground-state samples~\cite{Matsuda2009, Mandra2017, Konz2019}. Because such issues are often intrinsically tied to the hardware details underlying an Ising solver, a numerical study of sampling performance requires the development and study of accurate models for the machine and its operation.

In this paper, we study the sampling performance of the measurement-feedback-based coherent Ising machine (MFB-CIM)~\cite{McMahon2016, Inagaki2016, Inagaki2016b, Yamamoto2020}, a hardware platform originally conceived for performing Ising optimization using a network of degenerate optical parametric oscillators (DOPOs) subject to a real-time measurement-feedback protocol, which encodes Ising interactions into the network dynamics. In particular, we use a \emph{Gaussian-state model} to examine how quantum noise arises within the dynamics of the MFB-CIM and address whether stochastic nonlinear dynamics can facilitate efficient sampling of low-energy Ising configurations. While a full-quantum treatment of the MFB-CIM is possible~\cite{Yamamoto2020}, a numerical study of the large-scale systems relevant for combinatorial optimization/sampling is only possible up to the Gaussian-state regime where quantum correlations are considered up to second-order (i.e., up to covariances of observables)~\cite{Braunstein2005, Weedbrook2012}. This Gaussian-state approximation is consistent with the operational regimes of all experimental MFB-CIMs known to date~\cite{McMahon2016, Inagaki2016, Inagaki2016b}, while still providing an accurate treatment of important quantum-noise-driven phenomena such as squeezing/antisqueezing and measurement uncertainty and backaction~\cite{Yamamoto2020, Caves1982, Wiseman2010}, which are central to our study of sampling performance but usually neglected in mean-field models.

The potential of MFB-CIMs to generate samples of degenerate ground- and low-energy-excited spin configurations was recently pointed out in Ref.~\cite{Kako2020}, using a Gaussian-state model formulated in \emph{continuous time}~\cite{Inui2020}. As we show in this work, such continuous-time models correctly capture the dynamics of the MFB-CIM in the \emph{high-finesse} limit where the cavity decay time of its constituent DOPOs dominate all other system timescales. On the other hand, the intrinsically higher bandwidth of a \emph{low-finesse} system can, at least in principle, be leveraged to significantly reduce computational runtime; indeed, most experimental implementations of CIMs (both optically-coupled as well as measurement-feedback-based) utilize DOPOs operating in the low-finesse regime of short cavity decay times~\cite{Marandi2014b, McMahon2016, Inagaki2016, Inagaki2016b}. Low-finesse systems are more conveniently described in \emph{discrete time}, where dynamics occur via a sequence of discrete operations on the system state~\cite{Hamerly2016}. Theoretically, quantum treatments of MFB-CIMs in discrete time have been previously studied in Refs.~\cite{Clements2017, Yamamura2017}. While the latter study used a non-Gaussian model for the quantum state, which is only numerically tractable for small problem sizes, the former work indeed turned to a Gaussian-state model to study the linear dynamics of the MFB-CIM. In their Gaussian model, however, the nonlinear gain saturation---which can play an important dynamical role in the MFB-CIM near and above threshold---was only considered phenomenologically. To circumvent these limitations, we develop here a discrete-time Gaussian-state quantum model featuring a physical model for nonlinear gain saturation, allowing us to study low- and intermediate-finesse MFB-CIMs below, through, and above threshold. To our knowledge, the model presented here is the most general treatment currently available to numerically simulate large-scale MFB-CIMs operating in the Gaussian-state regime.

\section{Discrete-time Gaussian quantum model of the MFB-CIM}
\label{sec:discrete-time-model}

Conceptually, the coherent Ising machine (CIM) is a system of $N$ degenerate optical parametric oscillators (DOPOs), which are nonlinear optical oscillators exhibiting saturable phase-sensitive gain. When pumped below its oscillation threshold, the state of a DOPO is well described by a quadrature-squeezed vacuum state, while far above threshold, nonlinear saturation of the gain due to pump depletion stabilizes the system into one of two phase-bistable bright coherent states (referred to as the $0$- and $\pi$-phase states). To encode Ising spins into the DOPO network, we associate these bistable phase states to the Ising spins $\sigma = \pm 1$, respectively. By engineering the interactions among DOPOs, we can realize system dynamics dictated by a desired Ising coupling matrix $J$.

\begin{figure}[b]
    \includegraphics[width=0.48\textwidth]{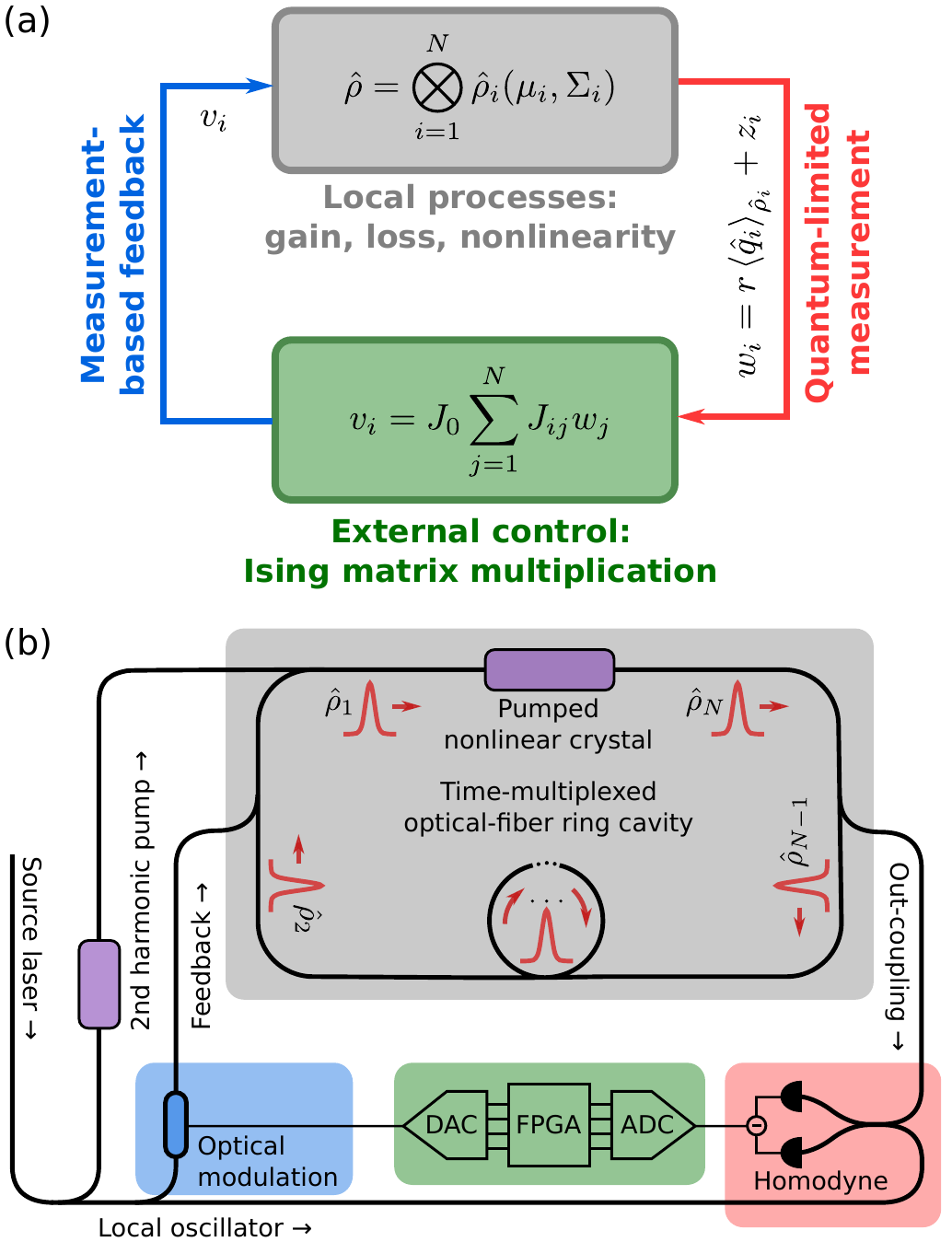}
    \caption{Schematic of the measurement-feedback coherent Ising machine (MFB-CIM). (a) Abstractly, the MFB-CIM consists of a system (gray box) of $N$ optical modes with state $\hat\rho_i$ (characterized by a mean vector $\mu_i$ and covariance matrix $\Sigma_i$), each of which experiences optical gain, loss, and nonlinearity. The system is probed via weak measurement to produce an estimate $w_i$ of the in-phase quadrature $\hat q_i$, up to some normally-distributed (i.e., Gaussian) quantum noise $z_i$. This estimate is processed by an external controller (green box) that generates a feedback signal $v_i$ based on a specified Ising-problem matrix $J_{ij}$. Closing the loop thus embeds the Ising couplings into the system dynamics. (b) An optical schematic of the MFB-CIM, implemented in a time-multiplexed scheme~\cite{McMahon2016,Inagaki2016}. The system consists of optical pulses $\hat\rho_i$ circulating in a resonant cavity with a pumped nonlinear crystal to provide gain and nonlinearity, plus any excess linear losses. An output coupler taps out a fraction of each pulse to be measured in balanced optical homodyne, producing the electronic signal $w_i$. This signal is processed by an FPGA to generate the feedback $v_i$, which is optically re-encoded by applying intensity/phase modulation to a local oscillator that is injected back into the cavity, optically displacing the pulses and closing the loop.}
    \label{fig:experimental-schematic}
\end{figure}

\begin{figure*}
    \includegraphics[width=0.96\textwidth]{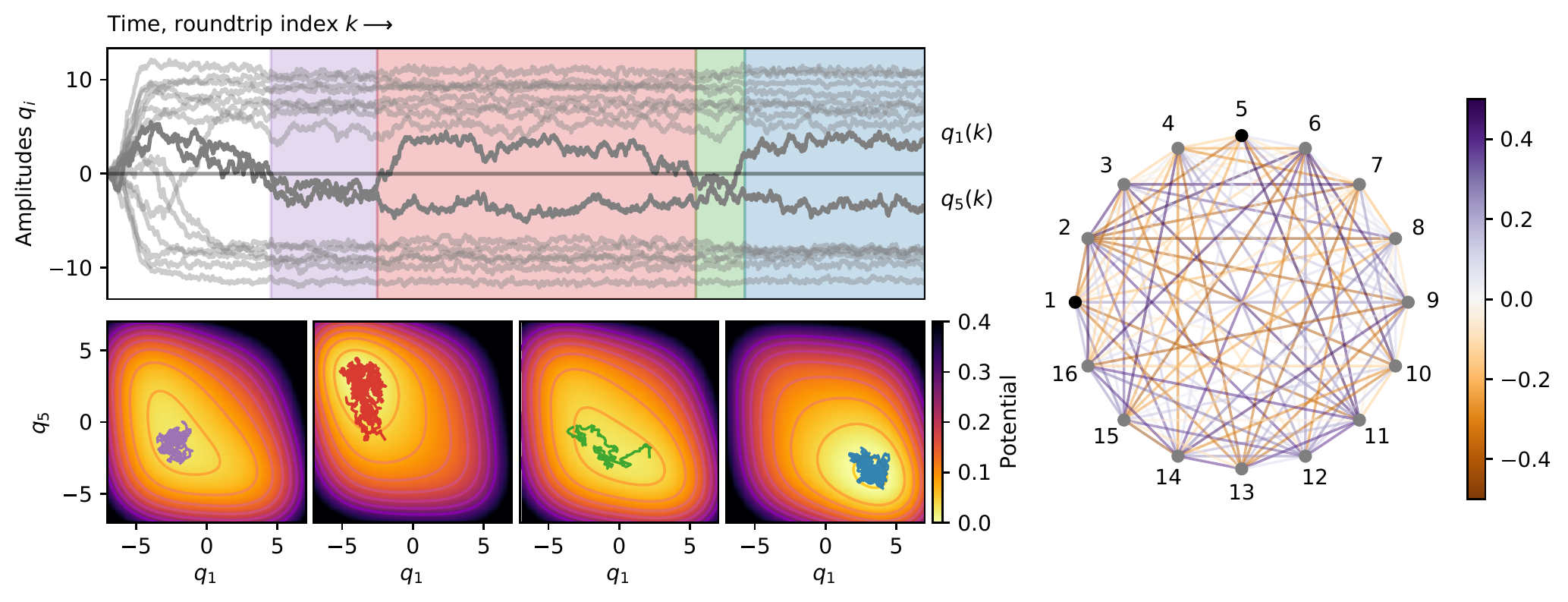}
    \vspace{-6pt}
    \caption{Conceptual illustration of the influence of quantum noise on the nonlinear dynamics of a MFB-CIM and the sampling of spin configurations from a representative Ising problem instance whose graph is shown to the right. The top-left panel shows the stochastic evolution of the $q$-quadrature expectation values $q_i$ of the $i$th pulse according to the discrete-time Gaussian model detailed in Sec.~\ref{sec:discrete-time-model} of this paper. In the continuous-time limit of the model, these dynamics can be intuitively seen as gradient descent on an $N$-dimensional potential; e.g., $\textstyle V(q) = -\sum_{i=1}^N \left[\frac12(p-\kappa-\gamma)q_i^2 - \frac18 g q_i^4 + \frac12\lambda \sum_{j=1}^N J_{ij} q_iq_j\right]$ via \eqref{eq:qdot-cont}. For each color-coded time window highlighted in the top panel, we plot in the bottom-left panels a 2-dimensional slice of the system trajectory in ($q_1$,$q_5$)-coordinates; to visualize the corresponding ($q_1,q_5$)-slice of $V(q)$, we average its instantaneous value as determined by all the other coordinates over the given time window. The sequence illustrates the random fluctuations in the state driven by quantum noise, causing $V(q)$ to stochastically guide the system state through different sign configurations, thus sampling various low-energy Ising configurations of the problem instance.}
    \label{fig:gradient-descent}
\end{figure*}

Figure~\ref{fig:experimental-schematic} depicts on the schematic level a notably successful experimental implementation of the CIM where the DOPOs are realized as synchronously-pumped, time-multiplexed ``signal'' pulses in a single optical fiber-loop cavity, with pulse interactions mediated by a synchronous, real-time measurement-feedback protocol. In this measurement-feedback-based CIM (MFB-CIM), the signal pulses are separated by a time interval $1/f_\text{rep}$; thus to fit $N$ pulses, the cavity length is $\sim Nc/f_\text{rep}$. On each roundtrip through the cavity, the signal pulses sequentially co-propagate through a nonlinear $\chi^{(2)}$ crystal alongside synchronous, externally injected pump pulses derived from the second harmonic of the source laser, which imparts phase-sensitive amplification along the in-phase $q$ quadrature. Next, the signal pulses are tapped out sequentially through an output coupler. The output is measured on a $q$-quadrature homodyne detector, which results in an indirect and weak measurement of the $q$-quadrature amplitude of the internal signal pulse. Crucially, the sign configuration of the $N$ homodyne measurements, say $(\sgn w_1, \ldots, \sgn w_N)$, constitutes a sampled Ising spin configuration under the correspondence $\sigma_i \leftrightarrow \sgn w_i$. Finally, to implement the interactions between the pulses, an FPGA receives the homodyne results and computes a feedback signal $v_i \propto \sum_{j=1}^N J_{ij} w_j$, which is applied to the corresponding $i$th signal pulse via synchronous external injection of a feedback pulse derived from the source laser but with intensity and phase determined by $v_i$ (e.g., using synchronous optical modulators). The interference between the injected pulse and the internal signal pulse steers the system towards lower-energy Ising spin configurations, thus dynamically realizing the Ising coupling matrix $J$ in the MFB-CIM system.

The result of embedding the structure of the Ising couplings into the system dynamics is that the evolution of the state is governed by the interplay among three general elements: (i) nonlinearity, which drives the signal amplitudes to bistable spin values; (ii) linear coupling, which drives the system towards collective configurations that minimize the Ising energy; and (iii) quantum noise, which arises from the inherent uncertainty of the weak homodyne measurement followed by measurement backaction and feedback injection and introduces stochasticity into the system dynamics. As illustrated conceptually in Fig.~\ref{fig:gradient-descent}, the dynamical evolution of the signal amplitudes is stochastic and nonlinear, but with a strong preference towards low-energy sign configurations dictated by the Ising coupling matrix. In the continuous-time limit, a convenient and intuitive picture is to think of such stochastic trajectories as quantum-noise-driven gradient descent on a potential landscape: As the state evolves stochastically in time, the instantaneous potential seen by any given spin dynamically changes as well, guiding the system around and across local minima born out of the interplay between nonlinearity and coupling. While we will not focus on elucidating this energy-landscape concept in this paper, such a multistability dynamical picture of MFB-CIM dynamics in the quantum-noise-dominated regime can provide useful physical intuition as we develop the formalism.

Open-dissipative bosonic systems like the MFB-CIM with relatively weak single-photon-induced nonlinearities and subject to continuous homodyne measurement can usually be well approximated by a Gaussian state, even as the system evolves through the classical bifurcation at threshold. Formally, the Gaussian-state approximation means the quantum state, conditioned on the measurement results, has a Wigner function well approximated by a Gaussian distribution; consequently, the state can be fully characterized by simply specifying a set of mean-field amplitudes and a set of covariances describing the quantum correlations and uncertainties of those amplitudes~\cite{Braunstein2005, Weedbrook2012}. Another very useful simplification (which applies more specifically to MFB-CIMs) is that the pulses, while interacting through measurement feedback, are nevertheless unentangled since the physics in Fig.~\ref{fig:experimental-schematic} only involve local operations and classical communication (LOCC) among the signal pulses, leading to zero covariance between different signal pulses.

Many of the operations involved in Fig.~\ref{fig:experimental-schematic}, including outcoupling, homodyne measurement, and feedback injection, are linear operations, and for Gaussian states, such operations have a particularly simple description if we use a discrete-time formalism for the dynamics, where we assume the signal pulses undergo a discrete transformation upon passing through each optical component. This discrete-time approach stands in contrast to more traditional continuous-time models in quantum optics, where the amplitudes evolve continuously in time under an effective system Hamiltonian and a set of Liouvillian superoperators representing continuous losses and measurements. Of course, the time-multiplexed, pulsed nature of the setup described in Fig.~\ref{fig:experimental-schematic} lends itself naturally to a discrete-time model when the pulse widths are short compared to their separation. A continuous-time model can be thought of as an appropriate approximation to the discrete-time model when the cavity finesse is high: In this limit, the single-roundtrip gain, loss, and measurement effects are all small, leading to small changes in the cavity state on every roundtrip, so the overall system dynamics are well described by coarse-grained, continuous-time differential equations. In order to study the impact of varying the cavity finesse (and hence the dynamical bandwidth of the machine) on sampling performance, however, we require the more general framework of a discrete-time formalism.

Crucially, the particular operation in the MFB-CIM that does not lend itself easily to a discrete-time Gaussian model is the propagation of the signal pulse through the crystal. Below threshold, this operation can be well described by linear phase-sensitive gain along the $q$ quadrature~\cite{Caves1982}, and this is modeled as a discrete-time squeezing operation in Ref.~\cite{Clements2017}. On the other hand, when the DOPO is near or above threshold, the co-propagating pump pulse can become depleted, which saturates the gain and leads to nonlinear dynamics and even physics beyond the Gaussian-state approximation in the presence of strong single-photon-induced quantum nonlinearities~\cite{Yamamura2017}. The main contribution of the model that follows is to prescribe a numerically efficient treatment of the gain saturation physics in the Gaussian-state regime, allowing the discrete-time Gaussian model to be extended through and above threshold while remaining consistent with continuous-time Gaussian-state models derived from standard quantum-optical models of the MFB-CIM in the high-finesse limit~\cite{Inui2020}.

In Sec.~\ref{sec:basic-formalism} we review the basic Gaussian-state formalism with which all the linear operations in the MFB-CIM can be succinctly described. In Sec.~\ref{sec:crystal-propagation} we derive the Gaussian equations of motion for nonlinear propagation through the crystal. We then summarize in Sec.~\ref{sec:discrete-map-recipe} the entire iterative procedure for propagating the state of the MFB-CIM through one roundtrip, which completes our discrete-time dynamical model. Finally, in Sec.~\ref{sec:cont-time-reduction}, we outline how our discrete-time model reduces to, and connects with, more conventional continuous-time models for the MFB-CIM.

\subsection{Basic formalism}
\label{sec:basic-formalism}

We abstract the time-multiplexed MFB-CIM as an $N$-mode bosonic system with mode annihilation operators $\hat a_i$ obeying $\Sbrak{\hat a_i, {\hat a_j}^\dagger} = \delta_{ij}$ and quadrature operators $\hat z \coloneqq (\hat q_1, \hat p_1, \ldots, \hat q_N, \hat p_N)$ defined so that $\sbrak{\hat z_k, \hat z_\ell} = \im\Omega_{k\ell}$ where $\Omega \coloneqq \bigoplus_{i=1}^N \begin{psmallmatrix}0&1\\-1&0\end{psmallmatrix}$ is the symplectic form. If the system is in a Gaussian state~\cite{Braunstein2005, Weedbrook2012, Adesso2014, Brask2021}, it is fully determined by only a mean vector $\mu$ and a covariance matrix $\Sigma$; i.e., the quantum state can be written as $\hrho{\mu,\Sigma}$, where its first-order moment (i.e., mean vector) is
\begin{subequations}
\label{eq:mean}
\begin{equation}
    \mu_k \coloneqq \tr\paren{\hat z_k \hat\rho},
\end{equation}
and its second-order moment (i.e., covariance matrix) is
\begin{equation}
    \Sigma_{k\ell} \coloneqq \tr\paren{\frac12\paren{\delta\hat z_k\,\delta\hat z_\ell + \delta\hat z_\ell\,\delta\hat z_k}\hat\rho},
\end{equation}
\end{subequations}
where $\delta\hat z \coloneqq \hat z - \mu$ is a vector of fluctuation operators for each quadrature. Because the MFB-CIM is additionally unentangled due to LOCC dynamics, we can apply the additional simplifications:
\begin{subequations}\label{eq:locc-gaussian}
\begin{equation}
    \mu = \bigoplus_{i=1}^N \mu^{(i)}
    \quad\text{and}\quad
    \Sigma = \bigoplus_{i=1}^N \Sigma^{(i)},
\end{equation}
where, explicitly,
\begin{align}
    \mu^{(i)} &\coloneqq \Paren{\vbrak{\hat q_i},\vbrak{\hat p_i}}, \\
    \Sigma^{(i)} &\coloneqq \begin{pmatrix}
    \vbrak{\delta\hat q_i^2} & \frac12\vbrak{\delta\hat q_i\delta\hat p_i + \delta\hat p_i\delta\hat q_i} \\
    \frac12\vbrak{\delta\hat q_i\delta\hat p_i + \delta\hat p_i\delta\hat q_i} & \vbrak{\delta\hat p_i^2}
    \end{pmatrix},
\end{align}
so that, instead of having $\mathcal O(N^2)$ entries in general, there are at most only $4N$ nonzero entries in the covariance matrix (and only $3N$ unique ones) for the MFB-CIM. Accordingly, the quantum state factorizes as
\begin{equation}
    \hrho{\mu,\Sigma} = \bigotimes_{i=1}^N \hrho{\mu^{(i)},\Sigma^{(i)}},
\end{equation}
\end{subequations}
as expected. Note that, here, for two vectors $\mu^{(1)}$ and $\mu^{(2)}$, $\mu^{(1)} \oplus \mu^{(2)}$ denotes their concatenation while for two matrices $\Sigma^{(1)}$ and $\Sigma^{(2)}$, $\Sigma^{(1)} \oplus \Sigma^{(2)}$ denotes the block diagonal matrix $\begin{psmallmatrix}\Sigma^{(1)}&0\\0&\Sigma^{(2)}\end{psmallmatrix}$.

More generally, when two systems with states $\hrho{\mu^{(a)}, \Sigma^{(a)}}$ and $\hrho{\mu^{(b)}, \Sigma^{(b)}}$ are brought together, the joint system is described by the state
\begin{align}
    &\hrho{\mu^{(a)}, \Sigma^{(a)}} \otimes \hrho{\mu^{(b)}, \Sigma^{(b)}} \nonumber\\
    &\qquad= \hrho{\mu^{(a)}\oplus\mu^{(b)}, \Sigma^{(a)} \oplus \Sigma^{(b)}}.
\end{align}
On the other hand, if $\hrho{\mu^{(a,b)},\Sigma^{(a,b)}}$ is a joint system of two modes, then we can partial trace out mode $b$ by projecting out the subspace associated with mode $b$:
\begin{subequations}
\begin{equation}
    \tr_b\sbrak{\hrho{\mu^{(a,b)},\Sigma^{(a,b)}}} \coloneqq \hrho{P\mu^{(a,b)}, P\Sigma^{(a,b)}P^\T},
\end{equation}
where the projection matrix in this case is
\begin{equation}
    P \coloneqq \begin{pmatrix}
    1 & 0 & 0 & 0 \\
    0 & 1 & 0 & 0
    \end{pmatrix}.
\end{equation}
\end{subequations}

Having established the basic formalism, we now briefly describe the linear operations that are necessary for the operation of the MFB-CIM before moving onto the nonlinear crystal propagation. These elementary operations consist of beamsplitters for modeling loss and outcoupling, coherent injections for modeling feedback, and homodyne measurements.

A two-mode beamsplitter acting on a two-mode state $\hrho{\mu,\Sigma}$ with field-exchange amplitude $r$ (i.e., power exchange ratio $r^2$) can be described as
\begin{subequations} \label{eq:beamsplitter-map}
\begin{equation}
    \beamsplitter_r\sbrak{\hrho{\mu,\Sigma}} \coloneqq \hrho{S\mu, S\Sigma S^\T}
\end{equation}
with the beamsplitter matrix
\begin{equation}
    S \coloneqq \begin{pmatrix}
    t & 0 & -r & 0 \\
    0 & t & 0 & -r \\
    r & 0 & t & 0 \\
    0 & r & 0 & t
    \end{pmatrix},
\end{equation}
\end{subequations}
where $t \coloneqq \sqrt{1-r^2}$ is the self-scattering amplitude.

A coherent injection of a displacement $\alpha \in \mathbb R^2$ (representing the two quadratures of the displacement) into a mode $a$ can be obtained by introducing a new mode $b$ with a displaced mean $\alpha/\varepsilon$ and then applying a beamsplitter with field-exchange amplitude $\varepsilon \rightarrow 0$ to $a$ and $b$. In this limit, the mode $a$ does not inject into $b$, but since the mean of $b$ goes as $\alpha/\varepsilon$, the overall displacement incurred by $a$ goes to a constant in the limit:
\begin{align*}
    &\displace_\alpha\sbrak{\hrho{\mu^{(a)},\Sigma^{(a)}}} \\
    &\quad{} \coloneqq \lim_{\varepsilon\rightarrow0} \tr_b\paren{\mathcal \beamsplitter_\epsilon\sbrak{\hrho{\mu^{(a)},\Sigma^{(a)}} \otimes \hrho{\varepsilon\inv\alpha^{(b)},\Sigma^{(b)}_0}}},
\end{align*}
where $\Sigma_0 \coloneqq \diag(1/2,1/2)$ is the covariance of a coherent state. The result of this limit is simple, and (dropping the superscripts for simplicity) gives the expected result
\begin{equation} \label{eq:displacement-map}
    \displace_\alpha\sbrak{\hrho{\mu,\Sigma}} = \hrho{\mu+\alpha,\Sigma}.
\end{equation}

Finally, we consider making a $q$-quadrature measurement of a mode $b$ in a two-mode system $\hrho{\mu^{(a,b)}, \Sigma^{(a,b)}}$, which we can write in the general form
\begin{equation*}
    \mu^{(a,b)} \coloneqq \begin{pmatrix}\mu^{(a)}\\\mu^{(b)}\end{pmatrix}
    \quad\text{and}\quad
    \Sigma^{(a,b)} \coloneqq \begin{pmatrix}
    \Sigma^{(a)} & V \\
    V^\T & \Sigma^{(b)}
    \end{pmatrix},
\end{equation*}
where $V$ captures the quantum correlation between the two modes. The measurement results in a random normally-distributed output
\begin{equation}\label{eq:homodyne-record}
    w \sim \mathcal N\Paren{\mu^{(b)}_q, \Sigma^{(b)}_{qq}},
\end{equation}
where $\mu^{(b)}_q$ and $\Sigma^{(b)}_{qq}$ ($q$ simply denotes the first index) are respectively the mean and variance of the $q$ quadrature of mode $b$. After the measurement is performed, the mode $b$ is projected onto a $\hat q_b$-eigenstate $\ket{q=w}_b$ and can be formally traced out. The appropriate backaction onto the mode $a$ is described by~\cite{Adesso2014, Brask2021}
\begin{align*}
    \mu_w^{(a)} &\coloneqq \mu^{(a)} + V\Paren{Q \Sigma^{(b)} Q}^+ \Paren{(w,0)^\T - \mu^{(b)}}, \\
    \Sigma_w^{(a)} &\coloneqq \Sigma^{(a)} - V\Paren{Q \Sigma^{(b)} Q}^+V^\T,
\end{align*}
where $Q \coloneqq \begin{psmallmatrix} 1 & 0 \\ 0 & 0\end{psmallmatrix}$ is the projector onto the $q$-quadrature of mode $b$ and for any matrix $M$, $M^+$ denotes its Moore-Penrose pseudo-inverse. That is, after obtaining the measurement result $w$, the conditional state of mode $a$ is $\hrho{\mu^{(a)}_w, \Sigma^{(a)}_w}$. An alternative, more explicit formula can be obtained for this simple two-mode case by writing $V = \begin{pmatrix} v_q & v_p\end{pmatrix}$. Then we can compute the pseudo-inverse analytically~\cite{Adesso2014, Brask2021} to get
\begin{subequations} \label{eq:homodyne-conditional-state}
\begin{align}
    \mu_w^{(a)} &= \mu^{(a)} + \paren{\frac{w-\mu^{(b)}_q}{\Sigma^{(b)}_{qq}}} v_q, \\
    \Sigma_w^{(a)} &= \Sigma^{(a)} - \frac{v_q v_q^\T}{\Sigma^{(b)}_{qq}}.
\end{align}
\end{subequations}
We denote the process of homodyne measurement plus backaction by the operation
\begin{equation} \label{eq:homodyne-map}
    \homodyne_b\sbrak{\hrho{\mu^{(a,b)},\Sigma^{(a,b)}}} \coloneqq \hrho{\mu_w^{(a)},\Sigma_w^{(a)}},
\end{equation}
conditional on the measurement output $w$ given by \eqref{eq:homodyne-record}.

While these linear maps describing outcoupling, measurement, and feedback injection are fairly straightforward, we also need to take into account dissipative linear losses as well. Experimental sources of loss in the physical CIM vary by implementation details, but some prominent sources include crystal facet losses (due to mode-matching inefficiency or Fresnel-reflection losses) and cavity propagation losses (due to scattering off mirrors or mode-matching inefficiency while coupling in and out of fibers). Since crystal facet losses generally dominate in realistic experimental implementations, we assume for simplicity that all loss mechanisms can be lumped together and applied via a pair of partial beamsplitters, placed before and after the crystal. Like the outcoupler used for measurement, these beamsplitters tap out intracavity light, but instead of the outgoing pulse being measured via homodyne (which would cause backaction on the state), we assume this external pulse cannot be measured and we simply partial trace it out instead, leading to dissipation on the state.

\subsection{Nonlinear crystal propagation}
\label{sec:crystal-propagation}

The most difficult part of the discrete-time model concerns the propagation of the pulse through the nonlinear crystal, which, as a dynamical non-Gaussian process, stands in contrast to the other operations, including measurement and feedback, that can all be ideally treated as Gaussian operations. We assume that for each $i$th incoming signal pulse in mode $\hat a_i$, a new pump pulse in mode $\hat b$ instantiated as a coherent state is injected into the optical path via a dichroic coupler to copropagate synchronously with the signal pulse, thus activating a parametric interaction between the signal and pump described by a Hamiltonian
\begin{equation}
\label{eq:crystal-hamiltonian}
    \hat H^{(i)}_\text{nl}/\hbar = \frac{\im \epsilon}{2} \paren{\hat b\hat a_i^{\dagger2} - \hat b^\dagger \hat a_i^2},
\end{equation}
where the coupling rate $\epsilon$ determines the overall small-signal parametric gain experienced by the signal pulse for a given crystal length and initial pump-pulse amplitude. The two-mode-interaction form of this Hamiltonian assumes that the pulses are either sufficiently long in time to avoid walk-off or other pulse distortion effects due to dispersion, or that such dispersion has otherwise been well managed, allowing us to abstract both the signal and pump pulses as single-mode excitations of the field. In such a model, mode-matching inefficiencies (temporal, spectral, spatial, etc.) are all taken into account by the coupling rate $\epsilon$.

In general, the Hamiltonian \eqref{eq:crystal-hamiltonian} can produce both entanglement and non-Gaussianity in the joint state between the pump and signal pulses, requiring the full joint Hilbert space of the two modes to describe properly. In order to make the crystal propagation compatible with the Gaussian formalism, we derive equations of motion (EOMs) for the Gaussian moments of the pump and signal pulses generated by \eqref{eq:crystal-hamiltonian}, while assuming that the non-Gaussianity of the state (characterized by higher-order moments) remains negligible. This approximation is valid if the DOPOs have a large saturation photon number, i.e., a single photon only induces small gain saturation. We can then numerically integrate the EOMs from the input to the output facets of the crystal, resulting in a nonlinear map, which we abstractly write as
\begin{equation}\label{eq:crystal-map-abstract}
    \chi: \hrho{\mu^{(i)},\Sigma^{(i)}} \otimes \hrho{\mu_0^{(b)},\Sigma_0^{(b)}} \mapsto \hrho{\mu^{(i,b)},\Sigma^{(i,b)}},
\end{equation}
which acts on the incoming state (a Gaussian signal pulse unentangled with a coherent-state pump pulse) and produces a correlated pump-signal Gaussian state. After the crystal propagation is complete, we need to also address what to do with the pump pulse, as it can, in general, be entangled with the signal. The option we take here is to trace out the pump mode, producing a mixed Gaussian state describing only the signal pulse; this state impurity of the signal pulse can be viewed as dissipation caused by two-photon absorption or, equivalently, energy loss due to back-conversion from signal to pump.

One straightforward way to restrict the quantum dynamics to a Gaussian subspace is to take the Heisenberg EOMs generated by \eqref{eq:crystal-hamiltonian} for the quadrature operators and perform a moment expansion up to second order~\cite{Vladimirov2012}. The Heisenberg EOMs for crystal propagation of the $i$th signal pulse $\hat a_i$ and its corresponding pump pulse $\hat b$ are
\begin{equation}
    \diff{\hat a_i}{\tau} = \epsilon\hat b\hat a_i^\dagger,
    \qquad\qquad
    \diff{\hat b}{\tau} = -\frac\epsilon2 \hat a_i^2.
\end{equation}
Let us now write for convenience $\hat a_i = \hat x_i + \im\hat y_i$ and $\hat b = \hat x\pump + \im\hat y\pump$, so that $\hat x_i \coloneqq \hat q_i/\sqrt2$ and $\hat y_i \coloneqq \hat p_i/\sqrt2$. Then these scaled quadrature operators evolve according to
\begin{align*}
    \diff{\hat x_i}{\tau} &= \epsilon\paren{\hat x\pump \hat x_i + \hat y\pump \hat y_i}, &
    \diff{\hat x\pump}{\tau} &= -\frac\epsilon 2 \paren{\hat x_i^2 - \hat y_i^2}, \\
    \diff{\hat y_i}{\tau} &= \epsilon\paren{\hat y\pump \hat x_i - \hat x\pump \hat y_i}, &
    \diff{\hat y\pump}{\tau} &= -\frac\epsilon2 \paren{\hat x_i\hat y_i + \hat y_i\hat x_i}.
\end{align*}

The evolution of the first-order moments can simply be obtained by taking expectation on the above equations. In order to break up the products, we can use the relation $\mean{\hat z_1 \hat z_2} = \covar{\hat z_1}{\hat z_2} + \mean{\hat z_1}\mean{\hat z_2}$ to express the expectation of a product of any two operators $\hat z_1$ and $\hat z_2$ in terms of their means and covariance. However, it is also clear that in doing so, we need to track the evolution of the covariances as well. To derive the covariance EOMs, we use the general formula
\begin{equation*}
    \diff{}{\tau}\covar{\hat z_1}{\hat z_2} = \mean{\diff{\hat z_1}{\tau}\hat z_2 + \hat z_1\diff{\hat z_2}{\tau}} - \diff{}{\tau}\Paren{\mean{\hat z_1}\mean{\hat z_2}}.
\end{equation*}
Crucially, in applying this equation, we make the Gaussian-moment assumption that
\begin{align}
    \mean{\hat z_1\hat z_2\hat z_3} &\mapsto \mean{\hat z_1}\covar{\hat z_2}{\hat z_3} + \mean{\hat z_2}\covar{\hat z_1}{\hat z_3} \\
    &\qquad{} + \mean{\hat z_3}\covar{\hat z_1}{\hat z_2} + \mean{\hat z_1}\mean{\hat z_2}\mean{\hat z_3},\nonumber
\end{align}
where the third-order (non-Gaussian) central moment $\mean{\delta\hat z_1\,\delta\hat z_2\,\delta\hat z_3} = 0$ by assumption.

The full EOMs derived under this procedure are provided in Appendix~\ref{app:full-eoms}. In general, since we can use $\sbrak{\delta\hat x, \delta\hat y} = \im/2$ to obtain $\covar{\hat y}{\hat x}$ from $\covar{\hat x}{\hat y}$, there are 10 covariances that we need to track. However, we can simplify the dynamics further by exploiting the properties of phase-sensitive amplification. Suppose that the initial state of the system obeys (i) $\mean{\hat y_i} = \mean{\hat y\pump} = 0$ (no quadrature-phase displacements) and (ii) $\mean{\{\delta\hat x_i,\delta\hat y_i\}} = \mean{\{\delta\hat x\pump,\delta\hat y\pump\}} = \covar{\hat x_b}{\hat y_i} = \covar{\hat y_b}{\hat x_i} = 0$ (all in-phase and quadrature-phase fluctuations are uncorrelated). We note linear loss and outcoupling are passive operations, which occur independently on the two quadratures, while the measurement and feedback injection act only on the $q$ quadrature, so none of the linear operations can produce a quadrature-phase displacement or generate correlations between the quadratures if none were there to begin with. For the crystal propagation, we can examine the full EOMs in Appendix~\ref{app:full-eoms}, which show that these conditions, if true at the input to the crystal, remain true throughout the crystal propagation. Thus we can take
\begin{subequations}
\label{eq:crystal-invariants}
\begin{align}
    \mean{\hat y_i} = \mean{\hat y\pump} &= 0, \\
    \mean{\textstyle\frac12\{\delta\hat x_i,\delta\hat y_i\}} = \mean{\textstyle\frac12\{\delta\hat x\pump,\delta\hat y\pump\}} &= 0, \\
    \covar{\hat x_b}{\hat y_i} = \covar{\hat y_b}{\hat x_i} &= 0,
\end{align}
\end{subequations}
to be invariants of the crystal propagation.

Following this procedure, we arrive at the final Gaussian-state EOMs, which can be numerically integrated to implement the crystal propagation map $\chi$ in \eqref{eq:crystal-map-abstract}. The mean equations are given by
\begin{subequations} \label{eq:crystal-eoms-mean}
\begin{align}
    \diff{\mean{\hat x_i}}{\tau} &= \epsilon\mean{\hat x\pump}\mean{\hat x_i} + \epsilon\mean{\delta\hat x\pump\,\delta\hat x_i + \delta\hat y\pump\,\delta\hat y_i}, \\
    \diff{\mean{\hat x\pump}}{\tau} &= -\frac\epsilon2 \mean{\hat x_i}^2 - \frac\epsilon2 \mean{\delta\hat x_i^2 - \delta\hat y_i^2},
\end{align}
\end{subequations}
while the covariance EOMs are
\begin{subequations} \label{eq:crystal-eom-variance}
\begin{align}
    \diff{\mean{\delta\hat x_i^2}}{\tau} &= + 2\epsilon\mean{\hat x\pump}\!\mean{\delta\hat x_i^2} + 2\epsilon\mean{\hat x_i}\!\covar{\hat x\pump}{\hat x_i}, \\
    \diff{\mean{\delta\hat y_i^2}}{\tau} &= - 2\epsilon \mean{\hat x\pump}\!\mean{\delta\hat y_i^2} + 2\epsilon\mean{\hat x_i}\!\covar{\hat y\pump}{\hat y_i}, \\
    \diff{\mean{\delta\hat x\pump^2}}{\tau} &= -2\epsilon\mean{\hat x_i}\!\covar{\hat x\pump}{\hat x_i}, \\
    \diff{\mean{\delta\hat y\pump^2}}{\tau} &= -2\epsilon\mean{\hat x_i}\!\covar{\hat y\pump}{\hat y_i}, \\
    \diff{\covar{\hat x\pump}{\hat x_i}}{\tau} &= \epsilon\mean{\hat x_i}\!\mean{\delta x\pump^2 - \delta x_i^2} + \epsilon\mean{\hat x\pump}\!\covar{\hat x\pump}{\hat x_i}, \\
    \diff{\covar{\hat y\pump}{\hat y_i}}{\tau} &= \epsilon\mean{\hat x_i}\!\mean{\delta\hat y\pump^2 - \delta y_i^2} - \epsilon\mean{\hat x\pump}\!\covar{\hat y\pump}{\hat y_i}.
\end{align}\end{subequations}
We may also explicitly specify the initial conditions for these EOMs. While $\mean{\hat x_i}$, $\mean{\delta\hat x_i^2}$, and $\mean{\delta \hat y_i^2}$ obviously depend on the state of the signal pulse input to the crystal, we have
\begin{subequations} \label{eq:ode-init-conditions}
\begin{align}
    \vbrak{\hat x_i}\!(0) &= \textstyle\frac1{\sqrt2}\vbrak{\hat q_i}, &
    \vbrak{\hat x\pump}\!(0) &= \textstyle\frac1{\sqrt2}\beta, \\
    \vbrak{\delta\hat x_i^2}\!(0) &= \textstyle\frac12 \vbrak{\delta\hat q_i^2}, &
    \vbrak{\delta\hat y_i^2}\!(0) &= \textstyle\frac12 \vbrak{\delta\hat p_i^2}, \\
    \vbrak{\delta\hat x\pump^2}\!(0) &= \textstyle\frac14, &
    \vbrak{\delta\hat y\pump^2}\!(0) &= \textstyle\frac14, \\
    \vbrak{\delta\hat x\pump\,\delta\hat x_i}\!(0) &= 0, &
    \vbrak{\delta\hat y\pump\,\delta\hat y_i}\!(0) &= 0,
\end{align}
\end{subequations}
where we have introduced $\beta \coloneqq \mean{\hat q_b}(0)$ as the $q$-quadrature displacement of the input coherent-state pump pulse; that is, $\beta/\sqrt{2}$ is its amplitude and $\beta^2/2$ is the expected photon number. Thus to implement the map \eqref{eq:crystal-map-abstract}, we integrate these initial conditions through the EOMs \eqref{eq:crystal-eoms-mean} and \eqref{eq:crystal-eom-variance} for a time $\tau_\text{nl}$, defined to be the time the pulse takes to propagate through the crystal. It is worth noting that the nonlinear coupling rate $\epsilon$ in \eqref{eq:crystal-hamiltonian} and the propagation time $\tau_\text{nl}$ only occur in our model as the dimensionless product $\epsilon\tau_\text{nl}$.

This system of ODEs consist of 8 real-valued dynamical variables and can be efficiently solved numerically; it is for this reason that we chose to use the quadrature operators $\hat x$ and $\hat y$ for this derivation, as the mode operators $\hat a$ and $\hat a^\dagger$ (which have complex-valued means and covariances) would have resulted in 8 complex-valued ODEs.

\subsection{Discrete-time dynamical model for the MFB-CIM}
\label{sec:discrete-map-recipe}

Having described all the components and transformations necessary to model the MFB-CIM, we now describe a concrete iterative procedure for generating the dynamics of the MFB-CIM. We let $\hrho{\mu^{(i)}(k),\Sigma^{(i)}(k)}$ denote the state of the $i$th pulse just before it starts its $k$th roundtrip through the system, which occurs at wall-clock time $(kN+i)/f_\text{rep}$, where $1/f_\text{rep}$ is the pulse repetition interval. Note that with this definition, the ``state'' $\bigotimes_{i=1}^N \hrho{\mu^{(i)}(k),\Sigma^{(i)}(k)}$ technically combines signal pulse states from different times, since pulse $i=1$ would have entered the next roundtrip (and possibly have already interacted with some optical elements) before pulse $i=N$ has finished the last roundtrip. Nevertheless, because the pulses experience LOCC evolution, this subtlety does not introduce a significant problem.

To propagate the state of the $i$th signal pulse from $\hrho{\mu^{(i)}(k),\Sigma^{(i)}(k)}$ to $\hrho{\mu^{(i)}(k+1),\Sigma^{(i)}(k+1)}$, we perform the following operations iteratively:
\begin{enumerate}
    \item \emph{Input facet loss}:  The input facet loss can be modeled as the operation
    \begin{align}\label{eq:facet-loss-map}
        &\quad\hrho{\mu^{(i)},\Sigma^{(i)}} \mapsto{} \nonumber\\
        &\qquad \tr_c \paren{\beamsplitter_{r_\text{loss}}\sbrak{\hrho{\mu^{(i)},\Sigma^{(i)}}\otimes\hrho{0^{(c)},\Sigma_0^{(c)}}}},
    \end{align}
    where $\mathcal B$ is the beamsplitter map defined by \eqref{eq:beamsplitter-map} and $r_\text{loss}^2$ is the power loss through that facet. Physically, $c$ represents a vacuum mode, which mixes with the signal pulse and is then traced out.
    \item \emph{Crystal propagation}: Following \eqref{eq:crystal-map-abstract}, the crystal propagation is described by a Gaussian map producing a joint correlated signal-pump state, followed by a partial trace of the pump mode:
    \begin{align}
        &\hrho{\mu^{(i)},\Sigma^{(i)}} \mapsto{}\nonumber\\
        &\quad \tr_b\sbrak{\chi\paren{\hrho{\mu^{(i)},\Sigma^{(i)}}\otimes\hrho{\beta^{(b)},\Sigma_0^{(b)}}}},
    \end{align}
    where the $b$ mode is a displaced coherent state with mean $\beta^{(b)} \coloneqq (\beta,0)$, and the map $\chi$ is obtained by solving the nonlinear Gaussian EOMs \eqref{eq:crystal-eoms-mean} and \eqref{eq:crystal-eom-variance} with initial conditions \eqref{eq:ode-init-conditions}. As described in Sec.~\ref{sec:crystal-propagation}, these EOMs involve the nonlinear interaction strength $\epsilon \tau_\text{nl}$ due to the action of the crystal Hamiltonian \eqref{eq:crystal-hamiltonian}, as well as the initial pump displacement $\beta \coloneqq \vbrak{\hat q\pump}\!(0)$.
    \item \emph{Output facet loss}: This step is exactly the same as for the input facet loss. Assuming we can lump the total system losses in a symmetric way between input and output losses around the crystal, we can again apply \eqref{eq:facet-loss-map}.
    \item \emph{Outcoupling and homodyne measurement}: The homodyne measurement consists of two steps. First, a part of the internal signal pulse is outcoupled, which can be described by the map
    \begin{subequations}
    \begin{align}
        &\quad\hrho{\mu^{(i)},\Sigma^{(i)}} \mapsto \hrho{\mu^{(i,h)},\Sigma^{(i,h)}} \nonumber\\
        &\qquad{} \coloneqq \beamsplitter_{r_\text{out}}\sbrak{\hrho{\mu^{(i)},\Sigma^{(i)}} \otimes \hrho{0^{(h)},\Sigma_0^{(h)}}},
    \end{align}
    where $r_\text{out}^2$ is the power outcoupling. This takes a probe external mode $h$ initialized in the vacuum state and mixes it with the signal pulse at the outcoupler to produce a correlated state of the internal cavity mode and the external outcoupled mode. The next step is to apply a homodyne measurement on the outcoupled mode, which produces a measurement result $w_i(k)$ for the $i$th signal pulse at this roundtrip index $k$ according to \eqref{eq:homodyne-record}. This indirect measurement of the internal signal pulse projects its state according to the map
    \begin{align}
        &\quad\hrho{\mu^{(i,h)},\Sigma^{(i,h)}} \mapsto \homodyne_h\sbrak{\hrho{\mu^{(i,h)},\Sigma^{(i,h)}}} \nonumber\\
        &{}\qquad= \hrho{\mu_{w_i}^{(i)},\Sigma_{w_i}^{(i)}}, 
    \end{align}
    \end{subequations}
    where $\homodyne$ is the conditional homodyne map \eqref{eq:homodyne-map}, with the mean and variances computed via \eqref{eq:homodyne-conditional-state}.
    \item \emph{Measurement-based feedback injection}: We finally apply displacements to the signal pulses based on the feedback signal computed by the FPGA for implementing the Ising couplings. Let the feedback terms be given by
    \begin{subequations}
    \label{eq:meas-feedback}
    \begin{equation}
        v_i(k) = J_0(k) \sum_{j=1}^N J_{ij} w_j(k),
    \end{equation}
    where $w_i(k)$ are the measurement results from the homodyne detection in this roundtrip, and $J_0(k)$ is a feedback gain parameter, which may generally depend on the roundtrip index $k$ (i.e., time). We now displace the pulse amplitudes according to
    \begin{equation}
        \hrho{\mu^{(i)},\Sigma^{(i)}} \mapsto \displace_{v_i} \sbrak{\hrho{\mu^{(i)},\Sigma^{(i)}}},
    \end{equation}
    \end{subequations}
    where $\displace$ is the displacement operation given by \eqref{eq:displacement-map}.
\end{enumerate}
The above steps, after being applied to each pulse $i = 1, \ldots, N$, completes one roundtrip through the CIM cavity. Note that the exact order in which we apply the above operations depends on the details of how the cavity is laid out and the relative time-of-flight between optical components, and our choice above is to some extent arbitrary. Nevertheless, generic features such as steady-state behavior should be robust against the exact choice of ordering, and if the precise transient behavior is desired (which can be important for very low-finesse operation), one can rearrange the procedure above to more accurately model the specific cavity layout.

\subsection{Reduction to continuous-time Gaussian models}
\label{sec:cont-time-reduction}

In this subsection, we briefly summarize how our discrete-time model can be reduced to continuous-time models more conventionally used in studies of optical CIMs in the ``high-finesse''. A complete derivation of this correspondence is presented in Appendix~\ref{app:cont-time}, and we only summarize the key ideas and results here.

In the high-finesse limit, each discrete operation only implements an infinitesimal change $\hat\rho \mapsto \hat\rho + \dif\hat\rho$ to the state $\hat\rho$ and, as in the Trotterization of quantum dynamics~\cite{Lloyd1996, Sakurai2017}, the exact order in which the operations are composed within one roundtrip becomes unimportant, allowing us to analyze the operations in Sec.~\ref{sec:discrete-map-recipe} independently within one roundtrip.

We introduce a parameter $\delta$ such that $\delta \rightarrow 0$ formally defines the high-finesse limit.  We begin with the assumption that the MFB-CIM roundtrip time (as measured by a wall clock) is $\Delta t = N/f_\text{rep} \sim \delta$. We then also assume that the model parameters in Sec.~\ref{sec:discrete-map-recipe} scale as follows:
\begin{align}
    \label{eq:cont-map-assump}
    r_\text{loss}^2 \sim r_\text{out}^2 \sim \beta^2 \sim (\epsilon \tau_\text{nl})^2 \sim J_0^2 \sim \delta.
\end{align}
As necessitated by working in the Gaussian regime, we also need to assume, for any fixed $\delta$, $(\epsilon \tau_\text{nl})^2 \ll r_\text{loss}^2 + r_\text{out}^2$.

In Appendix~\ref{app:cont-time}, we analyze each step of the discrete-map iteration from Sec.~\ref{sec:discrete-map-recipe} by expanding their effects on the Gaussian means and variances $\vbrak{\hat q_i}$ and $\vbrak{\delta\hat q_i^2}$ up to first order in $\delta$. Notably, the crystal propagation step can be similarly treated by using Picard iteration to integrate \eqref{eq:crystal-eoms-mean} and \eqref{eq:crystal-eom-variance} to first order in $\delta$, thus capturing the effects of parametric gain and pump depletion. After going through one entire roundtrip, we end up with updated means and variances with corrections up to first order in $\delta$. Denoting the updated state variables with a prime, the discrete-time map dynamics can then be connected to continuous-time derivatives via
\begin{subequations}
\label{eq:cont-disc-relation}
\begin{align}
    \lim_{\Delta t\rightarrow0} \frac{\vbrak{\hat q_i}' - \vbrak{\hat q_i}}{\Delta t} &\coloneqq \diff{\braket{\hat{q}_i}}{t}, \\
    \lim_{\Delta t\rightarrow0} \frac{\vbrak{\delta\hat q_i^2}' - \vbrak{\delta\hat q_i^2}}{\Delta t} &\coloneqq \diff{\braket{\delta\hat q_i^2}}{t}.
\end{align}
\end{subequations}
The continuous-time differential equations of motion have the explicit form
\begin{subequations} \label{eq:high-finesse-eoms}
\begin{align}
    \diff{\braket{\hat{q}_i}}{t} &= (p-\kappa-\gamma)\braket{\hat q_i} - \frac{g}{2}\braket{\hat q_i}^3 + \lambda \sum_{j=1}^N J_{ij} \braket{\hat q_j} \nonumber\\
    &\quad{} + 2\sqrt{\kappa}\paren{\braket{\delta\hat q^2_i} - \textstyle\frac{1}{2}} \xi_i + \frac{\lambda}{2\sqrt{\kappa}} \sum_{j=1}^N J_{ij} \xi_j \label{eq:qdot-cont} \\
    \diff{\vbrak{\delta\hat q_i^2}}{t} &= 2p\vbrak{\delta\hat q_i^2} - 2(\gamma+\kappa)\paren{\vbrak{\delta\hat q_i^2} - \textstyle\frac12} \\
    &\quad{}- 4\kappa\paren{\vbrak{\delta\hat q_i^2} - \textstyle\frac12}^2 - 2g\vbrak{\hat q_i}^2\paren{\textstyle\frac32\vbrak{\delta \hat q_i^2}-\frac12}. \nonumber
\end{align}
\end{subequations}
These equations are specified by the rates $\gamma$ (the intrinsic loss rate), $\kappa$ (the outcoupling rate), $p$ (the pump rate), $g$ (the nonlinear rate), and $\lambda$ (the feedback Ising-coupling rate), together with a set of white-noise processes $\xi_i$ obeying $\vbrak{\xi_i(t),\xi_j(t')} = \delta_{ij}\delta(t-t')$. In the limit $\Delta t \rightarrow 0$, they can be expressed in terms of discrete-time parameters as
\begin{subequations}
\label{eq:cont-disc-parameter-map}
\begin{align}
    \gamma &\coloneqq \frac{r_\text{loss}^2}{\Delta t}, &
    \kappa &\coloneqq \frac{r_\text{out}^2}{2 \Delta t}, \\
    p &\coloneqq \frac{\beta\epsilon \tau_\text{nl}}{\sqrt 2 \Delta t}, &
    g &\coloneqq \frac{(\epsilon \tau_\text{nl})^2}{4\Delta t}, \\
    \lambda &\coloneqq \frac{J_0 r_\text{out}}{\Delta t}, &
    \xi_i &\coloneqq \frac{z_i}{\sqrt{\Delta t}}.
\end{align}
\end{subequations}

As discussed in Appendix~\ref{app:cont-time}, continuous-time EOMs of the form \eqref{eq:high-finesse-eoms} have recently been shown to arise from quantum-optical master equations under appropriate Gaussian-state assumptions~\cite{Kako2020, Inui2020}, where the rates \eqref{eq:cont-disc-parameter-map} are the basic parameters in those models. Thus, while our model captures dynamics in the MFB-CIM beyond the high-finesse limit, it also reproduces the diffusive dynamics predicted by traditional quantum-optical models for the MFB-CIM in the appropriate limits. As such, Appendix~\ref{app:cont-time} may serve as a useful reference for readers interested in further exploring the relationship between continuous-time and discrete-time models of the MFB-CIM.

\section{Numerical results}
\label{sec:numerical_results}

In this section, we present and discuss small-scale numerical simulations of the discrete-time Gaussian model presented in Sec.~\ref{sec:discrete-time-model}.  We first show some representative trajectories of the model dynamics and define a suitable metric for sampling performance. We explore how sampling performance for a single problem instance depends on various model parameters, such as feedback gain or cavity finesse, and we verify that the sampling behavior is consistent across many different problem instances at small scale. Finally, we try to gain insight into the MFB-CIM dynamics by numerically studying a handful of operational modifications to the conventional MFB-CIM, such as the use of negative parametric gain, the removal of optical nonlinearity, and the replacement of quantum noise with classical noise.

\begin{figure*}
    \includegraphics[width=0.96\textwidth]{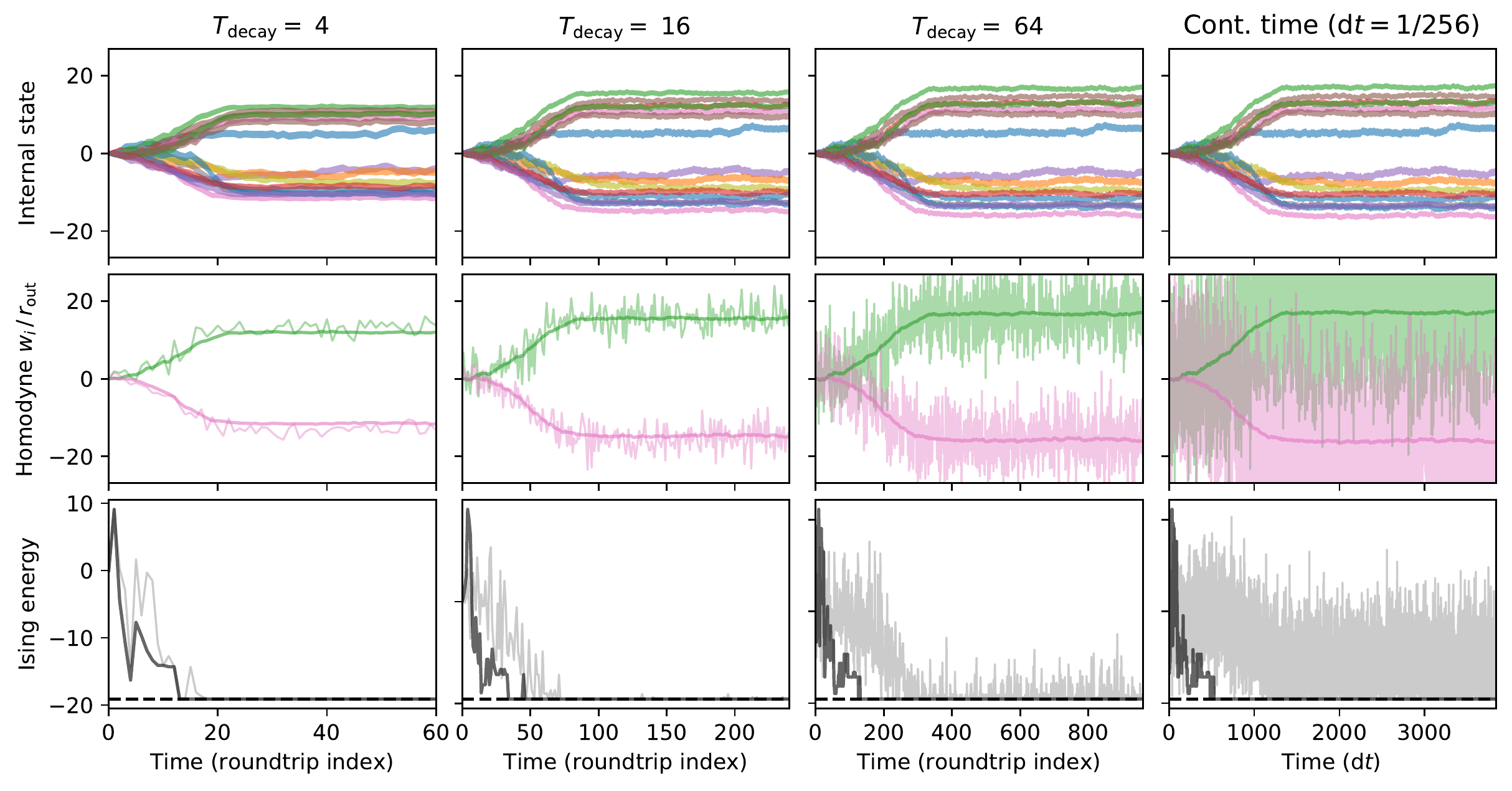}
    \caption{Representative trajectories of the discrete-time Gaussian-state MFB-CIM model for the $N = 16$ problem instance shown in Fig.~\ref{fig:gradient-descent}. Each of the first three columns represents a different cavity decay time $T_\text{decay}$, while the rightmost column depicts the continuous-time limit sampled at $\dif t = 1/256$; all trajectories are based on the same underlying noise process but sampled at different rates. The $N$ traces shown in the first row represent the Gaussian states of the $N$ intracavity signal pulses: the center of each trace gives the mean quadrature amplitude $\vbrak{\hat q_i}$, while the half-thickness of each trace is given by the root-variance $\vbrak{\delta\hat q_i^2}\!{}^{1/2}$. The second row shows two of the homodyne records $w_i$ (specifically $i = 2, 3$) divided by the outcoupling ratio $r_\text{out}$; the corresponding intracavity mean $\vbrak{\hat q_i}$ is reproduced from the first row as a guide for the eye. In the last row, the dark line shows the Ising energy $-\sum_j J_{ij} \sgn \vbrak{\hat q_i} \sgn \vbrak{\hat q_j}$ based on the intracavity mean, while the light-grey line shows the Ising energy $-\sum_j J_{ij} \sgn(w_i) \sgn(w_j)$ based on the homodyne record (black dashed line shows the ground energy). In this simulation, the other model parameters are held fixed at $n_\text{sat} = \num{200}$, $\eta_\text{esc} = \num{0.5}$, $r = \num{0.9}$, and $\alpha = \num{5}$.}
    \label{fig:dynamics}
\end{figure*}

\subsection{Model parameters}
For our numerical results, it is useful to define a new set of parameters, which scale the model more conveniently by keeping certain qualitative features of the dynamics constant. The dynamics of an uncoupled classical DOPO are critically determined by three parameters: (1) the cavity-photon $1/e^2$-decay time in the absence of pumping; (2) the pump parameter $r = \beta/\beta_\text{th}$ giving the ratio between the pump field $\beta$ over its threshold value $\beta_\text{th}$; and (3) the saturation photon number $n_\text{sat}$. We express each of these quantities in terms of the model parameters used in Sec.~\ref{sec:discrete-time-model}. For convenience, let us define
\begin{subequations} \begin{align}
    R_\text{out} &\coloneqq r_\text{out}^2, \\
    R_\text{loss} &\coloneqq 1 - \paren{1-r_\text{loss}^2}^2,
\end{align} \end{subequations}
where the latter quantity represents the total fraction of power lost through \emph{both} facets.

First, in the absence of pumping or nonlinearity, the number of roundtrips $T_\text{decay}$ required for the photon number to attenuate by a factor of $1/e^2$ due to linear loss and outcoupling is simply given by
\begin{equation} \label{eq:T-decay}
    1/T_\text{decay} \coloneqq -\log\sbrak{\paren{1-R_\text{out}}^{1/2}\paren{1-R_\text{loss}}^{1/2}}.
\end{equation}
In addition, because $T_\text{decay}$ captures the effect of $r_\text{loss}$ and $r_\text{out}$ together, it is also convenient to define an ``escape efficiency'' parameter
\begin{equation} \label{eq:eta-esc}
    \eta_\text{esc} \coloneqq \frac{R_\text{out}}{1-\paren{1-R_\text{out}}\paren{1-R_\text{loss}}},
\end{equation}
which captures the relative amount of (power) attenuation due to outcoupling as opposed to loss.

The threshold pump field is taken to be the value of $\beta$ (i.e., the $q$-quadrature displacement of the input pump pulse) such that the exponential gain experienced by a small-signal input to the crystal (i.e., a signal pulse with vanishing amplitude) exactly balances the attenuation due to linear loss and outcoupling. The pump parameter is then simply the pump field divided by this threshold value $\beta_\text{th}$. We therefore define
\begin{equation} \label{eq:beta_th}
    r \coloneqq \frac{\beta}{\beta_\text{th}},
    \quad\text{where}\quad
    \beta_\text{th} \coloneqq \frac{\sqrt2}{\epsilon \tau_\text{nl}} \frac{1}{T_\text{decay}}.
\end{equation}

The saturation photon number is the steady-state photon number at $r = 2$, considering the effects of loss and outcoupling, parametric gain, and nonlinear gain saturation. Because this feature involves the nonlinear terms of the crystal EOMs at finite signal amplitude, the exact value of the saturation photon number can depend on cavity layout for low-finesse cavities. In the high-finesse limit, however, it can be shown to be
\begin{equation} \label{eq:n_sat}
    n_\text{sat} = \frac{8}{(\epsilon \tau_\text{nl})^2} \frac{1}{T_\text{decay}}.
\end{equation}
When the roundtrip attenuation is moderately low ($\sim \num{0.4}$ in power), then $\epsilon \tau_\text{nl} \ll 1$, and we can take for convenience the above equation to \emph{define} the parameter $n_\text{sat}$, so that for a fixed $T_\text{decay}$, specifying $n_\text{sat}$ determines $\epsilon \tau_\text{nl}$, which then fixes $\beta_\text{th}$. When the roundtrip attenuation is large, however, it may be the case that a given $n_\text{sat}$ corresponds to $\epsilon \tau_\text{nl} \not\ll 1$, which is inconsistent with a Gaussian-state approximation for the crystal propagation. To handle these cases as well, we specify
\begin{equation} \label{eq:n_sat2}
    (\epsilon \tau_\text{nl})^2 = \min\paren{\frac{8}{n_\text{sat}T_\text{decay}}, \num{e-2}},
\end{equation}
where \num{e-2} is taken as an appropriate maximal value to respect the Gaussian-state approximation. In this latter case, \eqref{eq:beta_th} and \eqref{eq:n_sat} are replaced with $\beta_\text{th} = (10\sqrt2)/T_\text{decay}$ and $n_\text{sat} = 800/T_\text{decay}$.

In summary, we henceforth parametrize our system using the values $T_\text{decay}$, $\eta_\text{esc}$, $r$, and $n_\text{sat}$. We use \eqref{eq:T-decay} and \eqref{eq:eta-esc} to determine $r_\text{loss}$ and $r_\text{out}$, and we use \eqref{eq:n_sat2} to determine $\epsilon\tau_\text{nl}$. This procedure sets $\beta_\text{th}$ via \eqref{eq:beta_th}, which also gives us $\beta$ given $r$.

Finally, with regards to the feedback control, we note that for $r < 1$, the feedback gain $J_0$ needed for the system to go above threshold due to feedback gain scales with $\sqrt{T_\text{decay}}$ but also with the Ising matrix entries $J_{ij}$. To this end, we define a feedback gain parameter
\begin{equation}
    \alpha \coloneqq -J_0 \sqrt{T_\text{decay}} \paren{\textstyle \sum_{i \neq j} |J_{ij}|}^{1/2},
\end{equation}
where the negative sign is chosen since we usually use $J_0 < 0$ in order for the feedback to enforce minimization of the Ising energy; thus $\alpha$ is positive in these cases.

In Fig.~\ref{fig:dynamics}, we illustrate some representative dynamics of the MFB-CIM running on the $N=16$ problem instance shown in Fig.~\ref{fig:gradient-descent}. As a way of making the discussion in Sec.~\ref{sec:cont-time-reduction} more concrete, we note in particular how these trajectories change as a function of $T_\text{decay}$ while all other model parameters are held constant. By running the simulation for $15 T_\text{decay}$ roundtrips in all cases, we see there is a qualitative difference in going from a low-finesse system ($T_\text{decay} = 4$), to an intermediate-finesse one ($T_\text{decay} = 16$), to a high-finesse one ($T_\text{decay} = 64$), but the dynamics eventually converge to the continuous-time trajectory in the high-finesse limit, as depicted in the rightmost column. As expected, the homodyne record $w_i$ (and hence the measured Ising energy $-\sum_{j=1}^N J_{ij} \sgn w_i \sgn w_j$) becomes increasingly noisy as the finesse increases because the outcoupling ratio $r_\text{out} \sim 1/\sqrt{T_\text{decay}}$, providing less information about the internal state in any given shot of the measurement and requiring more roundtrips to obtain the same amount of information produced by a single measurement shot in a lower-finesse system. In addition, if we assume the wall-clock roundtrip time is fixed (corresponding to the time between successive points in the discrete-time model, or to $\dif t$ in the continuous-time model), it follows that the low-finesse system takes a shorter wall-clock time to reach steady state (i.e., $\sim 40$ roundtrips at $T_\text{decay} = 4$ vs $\sim 640$ roundtrips at $T_\text{decay} = 64$), all else being equal. This scaling plays an important role in the efficiency of the system and the overall time-to-sample as we analyze next.

\subsection{Ising sampling in Gaussian MFB-CIMs}
As evident from Fig.~\ref{fig:dynamics}, the dynamics of the MFB-CIM drive the system towards states encoding low-energy spin configurations of the Ising problem. At the same time, the particular configurations found by the MFB-CIM are stochastic. We therefore expect that, at least in certain regimes of operation, the MFB-CIM can be used to \emph{stochastically sample} different spin configurations, simply by running the system under the injection of measurement noise. Each ``run'' of the MFB-CIM would consist of a homodyne record like the ones shown in Fig.~\ref{fig:dynamics}, which takes $\mathcal O(T_\text{decay})$ roundtrips to collect and would yield one or more samples of low-energy Ising spin configurations after an initial transient period; repeated runs (i.e., passing through threshold again) could also generate new samples. The sampling efficiency is thus characterized by the likelihood of a given trajectory to yield at least one sample of interest and also how quickly it can do so, while sampling fairness depends on how uniformly such configurations are distributed.

\begin{figure}[h!b]
    \includegraphics[width=0.48\textwidth]{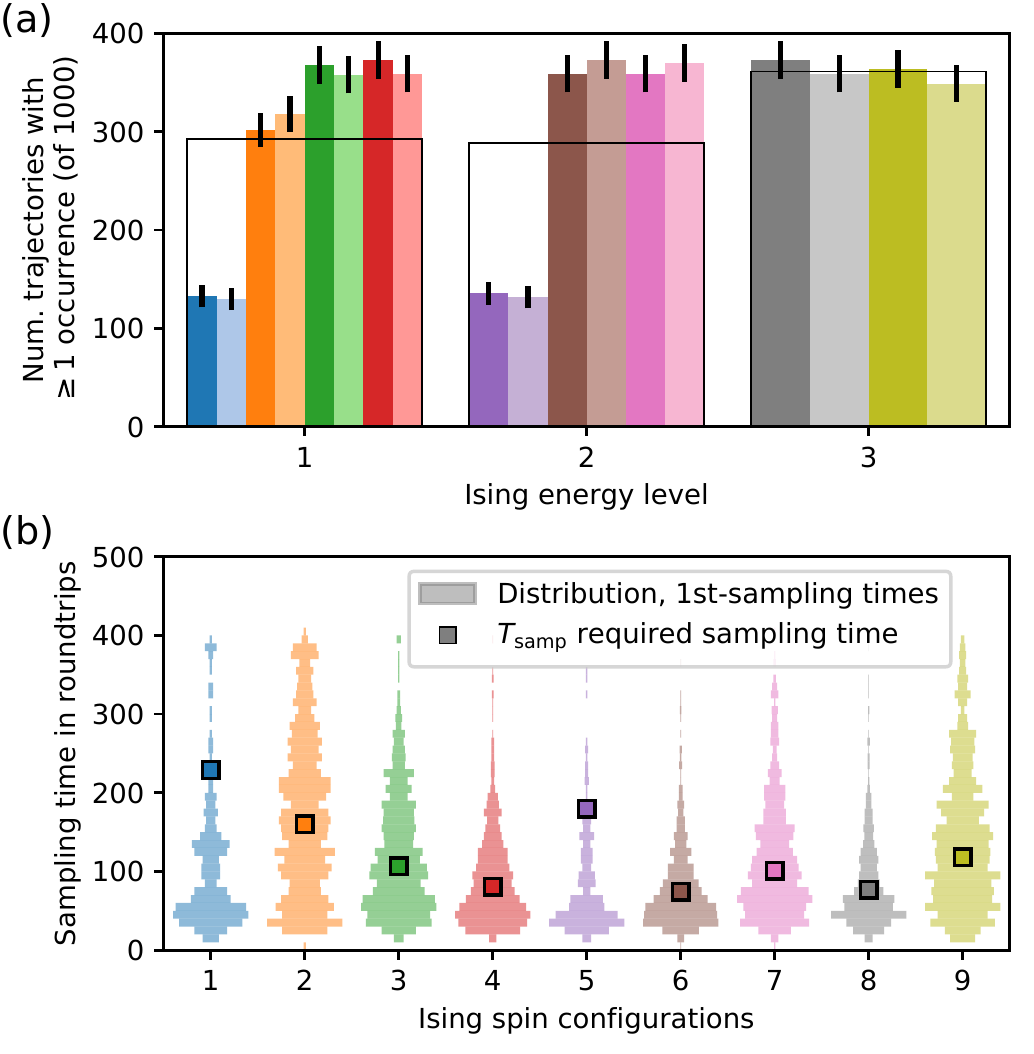} \hfill \vspace{-12pt}
    \caption{Sampling low-energy spin configurations of the $N=16$ Ising problem shown in Fig.~\ref{fig:gradient-descent} using an MFB-CIM. (a) Bar graph showing the number of trajectories with at least one occurrence of each spin configuration, each represented by a colored bar (adjacent bars of the same color but different lightness represent configurations differing only in an overall sign flip). Configurations are grouped together by Ising energy, with labels 1, 2, and 3 for the ground, first-excited, and second-excited energy levels, respectively. The solid-line rectangle around each group represents the expected histogram if sampling within that energy level were perfectly fair. Error bars indicate simulation uncertainty of the counts. (b) Resolving the time along a trajectory at which a sample first appears. The vertical histograms show, for the set of realized samples, the distribution of time (in roundtrips) at which the first sample appeared (i.e., the first-sampling time). Each spin configuration is colored in accordance with (a); however, occurrences of configurations differing only in an overall sign flip have been combined. Square markers show the required sampling time based on $T_\text{samp}$ as defined by \eqref{eq:req-samp-time}, a metric taking into account both the first-sampling time distribution as well as the frequency of occurrences. For this simulation, $r = \num{0.8}$, $\alpha = \num{5}$, $T_\text{decay} = \num{4}$, $\eta_\text{esc} = \num{0.2}$, and $n_\text{sat} = \num{200}$; $N_\text{traj} = \num{1000}$ trajectories are simulated for $\num{100} T_\text{decay}$ roundtrips each.}
    \label{fig:sampling-metric}
\end{figure}

Figure~\ref{fig:sampling-metric} illustrates this procedure for the $N=16$ problem instance shown in Fig.~\ref{fig:gradient-descent}. In Fig.~\ref{fig:sampling-metric}(a), the number of trajectories where each spin configuration appears at least once is recorded. Over the course of \num{1000} trajectories, we easily obtain multiple samples of every spin configuration of interest, indicating relatively fair sampling at least for this instance. Nevertheless, there are systematic biases in the sampling: namely, the spin configurations in a given energy level are not necessarily uniformly sampled. These biases are problem-dependent in general but also depend on the model parameters chosen for the sampling process.

To fully quantify the sampling efficiency, however, it is not enough to simply count trajectories in which each spin configuration appears, since certain configurations may systematically appear later than other configurations within any given trajectory. These differences in sampling time are illustrated in Fig.~\ref{fig:sampling-metric}(b), where, for each spin configuration considered in Fig.~\ref{fig:sampling-metric}(a) (up to an overall sign flip), we show a (vertical) histogram of the \emph{first time} that configuration appeared in each trajectory, if it appeared at all. There is an initial transient period ($\sim T_\text{decay}$ roundtrips) in which low-energy samples cannot be generated, and generally most of the distributions are peaked within a few decay times of the transient. However, the exact distribution of this first-sampling time differs across spin configurations, with some featuring sharper peaks and others having longer tails. As a result, a configuration appearing less often on a per-trajectory basis may nevertheless be efficient to sample if it tends to appear earlier.

We can define a ``required sampling time'' metric in order to take into account these effects, including the biases in overall counts, the transient-time costs, and the variation in the first-sampling-time distributions. Let us suppose we have collected an ensemble of homodyne records $w_i^{(\ell)}(k)$, where $1 \leq \ell \leq N_\text{traj}$ denotes different trajectories, $1 \leq i \leq N$ denotes the DOPO or spin index, and $k \geq 1$ indexes the number of roundtrips elapsed. Suppose further we are interested in sampling a particular Ising spin configuration $\sigma \coloneqq (\sigma_1, \ldots, \sigma_N)$. Then we define the first-sampling time of $\sigma$ in trajectory $\ell$ as
\begin{subequations} \label{eq:req-samp-time}
\begin{equation}
    T^{(\ell)}_\text{samp}(\sigma) \coloneqq \min_{k\geq0} \left\{k : \forall i\leq N,\, \sgn\Paren{w_i^{(\ell)}(k)} = \sigma_i\right\},
\end{equation}
where we take $\min\varnothing = \infty$ by convention for trajectories that produce no samples of $\sigma$. Then given a sufficiently large number of trajectories $N_\text{traj}$, each simulated for a sufficiently long time, an estimate for the required number of roundtrips to sample $\sigma$ is $T_\text{samp}(\sigma)$, where
\begin{equation}
    \frac{1}{T_\text{samp}(\sigma)} \coloneqq \frac{1}{N_\text{traj}} \sum_{\ell = 1}^{N_\text{traj}} \frac{1}{T_\text{samp}^{(\ell)}(\sigma)}.
\end{equation}
\end{subequations}
Under this metric, a spin configuration that is realized less often (so $1/T_\text{samp}^{(\ell)} = 0$ for more values of $\ell$) will have a larger $T_\text{samp}$, as will a spin configuration that takes longer to appear (so $T_\text{samp}^{(\ell)}$ is finite but large). As a result, $T_\text{samp}(\sigma)$ captures, in an \textit{a posteriori} sense, the observed efficiency for sampling the configuration $\sigma$.

\begin{figure}[b]
    \centering
    \includegraphics[width=0.5\textwidth]{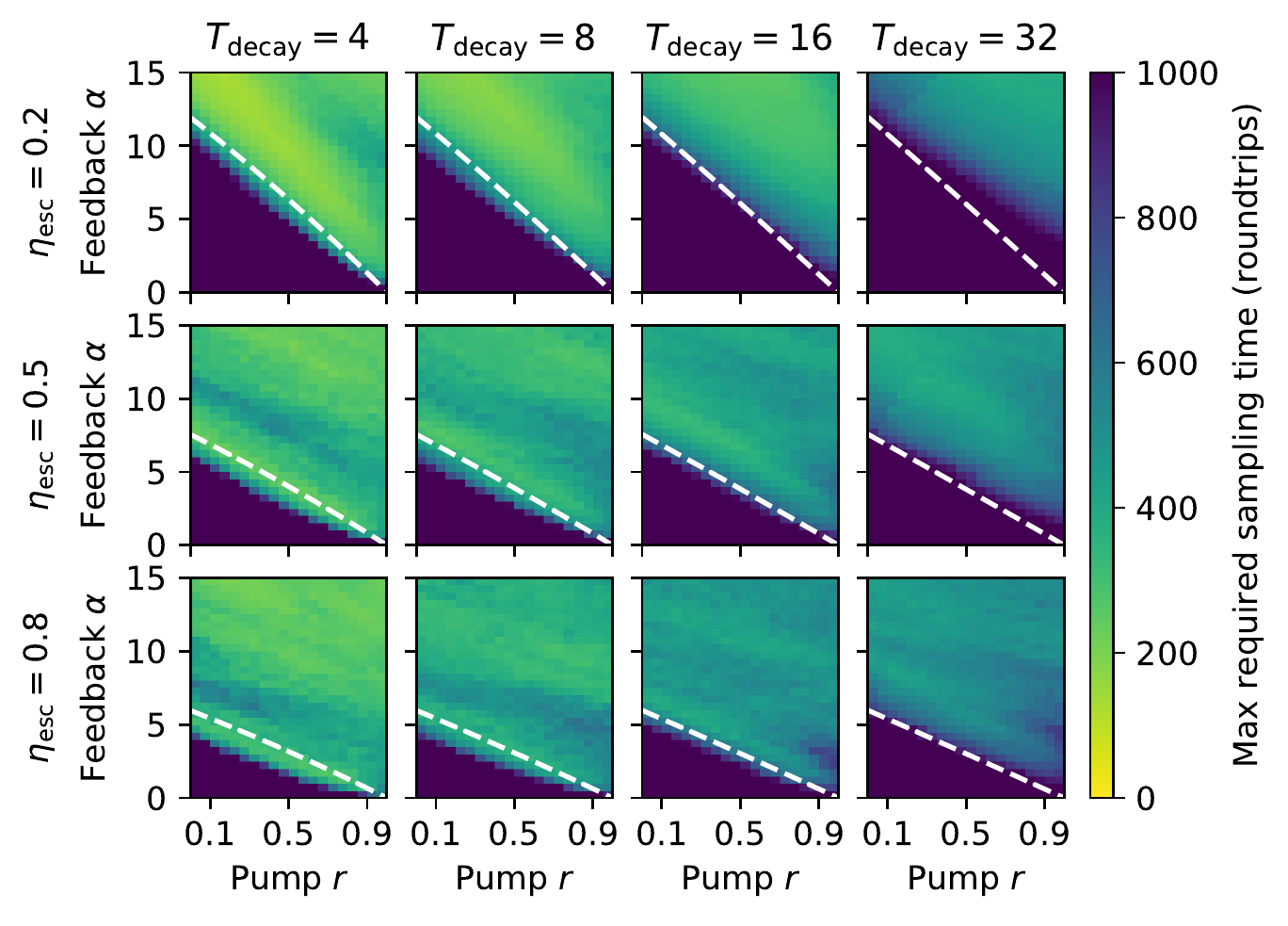}
    \caption{Required sampling time $T_\text{samp}$ in roundtrips as a function of model parameters for the problem instance and spin configurations considered in Fig.~\ref{fig:sampling-metric}. For each choice of model parameters, we run $N_\text{traj} = \num{1000}$ trajectories for $100 T_\text{decay}$ roundtrips and compute $T_\text{samp}(\sigma)$ according to \eqref{eq:req-samp-time} for each spin configuration $\sigma$, taking the maximum over the 9 configurations. As in Fig.~\ref{fig:sampling-metric}(b), we combine samples from configurations differing only in an overall sign flip. The dashed-white line indicates the threshold of the system, defined to be the boundary in $\alpha$ and $r$, beyond which there exists a nonzero-amplitude fixed point of the system dynamics at steady state. In all these simulations, we fix $n_\text{sat} = \num{200}$.}
    \label{fig:sampling-param-map}
\end{figure}

\begin{figure*}[t]
    \centering
    \includegraphics[width=0.92\textwidth]{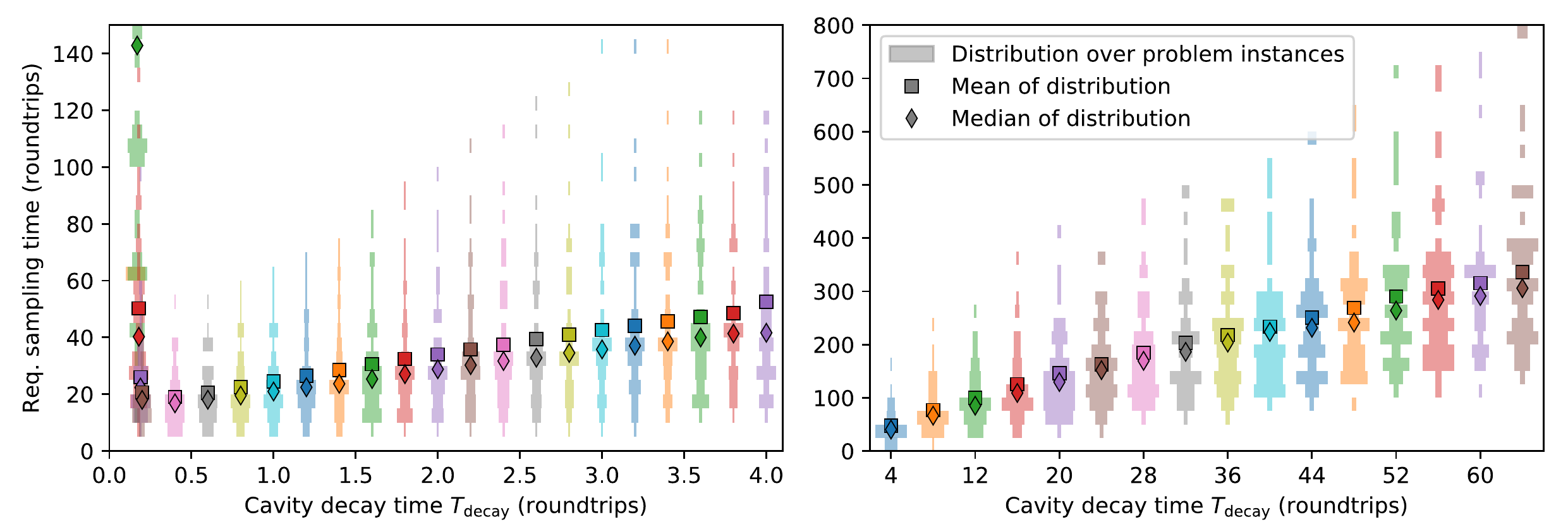}
    \caption{Time $T_\text{samp}(\sigma)$ required to sample a specified Ising ground-state configuration $\sigma$ as a function of the cavity decay time $T_\text{decay}$ for a set of 50 $N=16$ SK1 problems. For concreteness, we choose $\sigma$ as the first lexicographic ground-state configuration each problem (up to an overall sign flip). Each vertical histogram shows the distribution of $T_\text{samp}(\sigma)$ as defined by \eqref{eq:req-samp-time} over the different problem instances. Diamond markers indicate the median of the distribution while squares indicate the mean. Note that when the required sampling time becomes large, the mean may not be defined due to some problem instances requiring more trajectories to sample than were performed. For these simulations, we take $r = 0.8$, $\alpha = 4$, $\eta_\text{esc} = 0.2$, and $n_\text{sat} = 200$; we run $N_\text{traj} = 1000$ trajectories for $100 T_\text{decay}$ roundtrips.}
    \label{fig:finesse-scaling}
\end{figure*}

Having defined this empirical measure of required sampling time, we can turn to how it is affected by the model parameters, especially the feedback gain $\alpha$ and the pump parameter $r$. Figure~\ref{fig:sampling-param-map} shows how, for the 9 configurations considered in Fig.~\ref{fig:sampling-metric}, the largest required sampling time $T_\text{samp}$ varies with $\alpha$ and $r$. Efficient sampling in the MFB-CIM is relatively robust across a wide range of system parameters. The most critical parameters are indeed the feedback gain and the pump parameter, which show a sharp cutoff in sampling performance near the estimated linear threshold of the MFB-CIM. The required sampling time is lower for systems with a faster cavity decay time (i.e., $T_\text{decay} = 4$), which reflects the fact that a low-finesse cavity spends fewer roundtrips in the transient period and can yield low-energy samples more quickly due to larger (i.e., nondiffusive) kicks from the noise in each step. There appears to be a slight advantage to using a system with lower escape efficiency (higher background losses), which may be related to the fact that background loss affects the dynamical correlation between the pulse amplitudes differently from the noise due to measurement outcoupling~\cite{Yamamoto2020,Inui2020}.

One particularly important aspect of the sampling behavior in the MFB-CIM is that the required sampling time scales with the finesse of the system as measured by $T_\text{decay}$. To check whether this scaling is robust with respect to the choice of problem instance, we consider a set of integer-valued Sherrington-Kirkpatrick spin-glass Ising problems with range 1 (SK1), which is equivalent to a set of MAX-CUT problems with binary-signed edge weights. In Fig.~\ref{fig:finesse-scaling}, we show the distribution of $T_\text{samp}(\sigma)$ over 50 SK1 $N=16$ problem instances, where, for concreteness, $\sigma$ is chosen to be the first lexicographic spin configuration that gives the ground energy of the problem. We see that although there is a spread in the required sampling time across problem instances, the distributions are characterized by means and medians, which show a clear monotonic decrease with decreasing decay time. Interestingly, this scaling persists to very low decay times on the order of $T_\text{decay} \sim 1$, which is well outside the validity of any high-finesse or continuous-time model.  In fact, for this problem size, performance only saturates and degrades at $T_\text{decay} \approx 0.2$, which, for this parameter set, is the point at which the roundtrip attenuation begins to exponentially approach unity (i.e., $R_\text{loss} \sim 1-\e{-c/T_\text{decay}}$) as a function of $T_\text{decay}$. At this point, the sensitivity of the system to system parameters precludes any additional significant gains in reducing the sampling time. The fact that the sampling performance continues to improve into the low-finesse regime is a key motivation for the development of our discrete-time Gaussian-state model.

\subsection{Sampling in alternative models of MFB-CIM}
\label{sec:alternative-models}
Although our focus thus far has been on developing a general model for the MFB-CIM in the Gaussian-state approximation and studying the dynamical role of quantum noise in its conventional operation (with parametric gain, homodyne measurement/feedback, measurement backaction, and gain saturation), it is also useful to consider alternative models or modes of operation, which may be conceptually simpler or easier to implement experimentally. Of particular interest is to relate our quantum-based model to established classical analogs or formulations of CIMs, such as those based on coherent-state feedback networks without nonlinearity~\cite{Clements2017}, or those based on deterministic nonlinear dynamics (with no quantum noise and only a random initial condition), which have proven to be fruitful models in which to study the roles of feedback and nonlinearity for CIM combinatorial optimization~\cite{Wang2013, Leleu2019, Strinati2020}.

In this subsection, we consider three cases:
\begin{enumerate}
    \item \emph{MFB-CIM with zero or negative parametric gain}: Conventionally, the MFB-CIM is operated with parametric gain, i.e., the pump parameter $r > 0$. However, we can also explore its sampling performance for $r \leq 0$, with the case of $r=0$ being especially experimentally interesting as it does not require a pump source. Such modifications are straightforward within our general Gaussian model, so we can directly compare these cases against the conventional $r > 0$ case while keeping gain saturation, quantum noise, and so on fixed.
    \item \emph{MFB-CIM without nonlinear crystal}: We can also go one step further and consider a MFB-CIM without any parametric interaction (i.e., no optical nonlinearity) by setting $\epsilon \tau_\text{nl} = 0$, resulting in a ``coherent-state'' MFB-CIM, where the internal field is only excited through external coherent-state injection. This model has previously been studied in Ref.~\cite{Clements2017} in the context of combinatorial optimization (also via a discrete-time formulation), whereas we investigate here its potential for sampling. Since the resulting system has linear dynamics, the Gaussian formalism applies exactly and is an efficient representation of the quantum state throughout the dynamics.
    \item \emph{Mean-field MFB-CIM with injected measurement noise}: A common approach to studying open-dissipative optical systems with weak single-photon nonlinearities is to neglect quantum noise altogether by taking a mean-field or classical limit, resulting in deterministic c-number EOMs. We describe how such a limit can be taken for our Gaussian MFB-CIM model, producing not only the usual continuous-time mean-field models for the MFB-CIM~\cite{Wang2013, Leleu2019} but also a discrete-time mean-field model similar to that of Ref.~\cite{Hamerly2016} as well. However, to study sampling performance in this limit, we need an alternative noise source in the mean-field model. For this purpose, we supplement the model by injecting fixed-variance Gaussian-distributed noise (limiting to white noise at the continuous-time limit) in the measurement-and-feedback step~\cite{Pierangeli2020}; such an extrinsic noise source can correspond, for example, to classical Johnson noise in the detector or to a random signal intentionally generated by the FPGA circuit (e.g., via a pseudorandom number generator).
\end{enumerate}

\begin{figure}[hb!]
    \centering
    \includegraphics[width=0.47\textwidth]{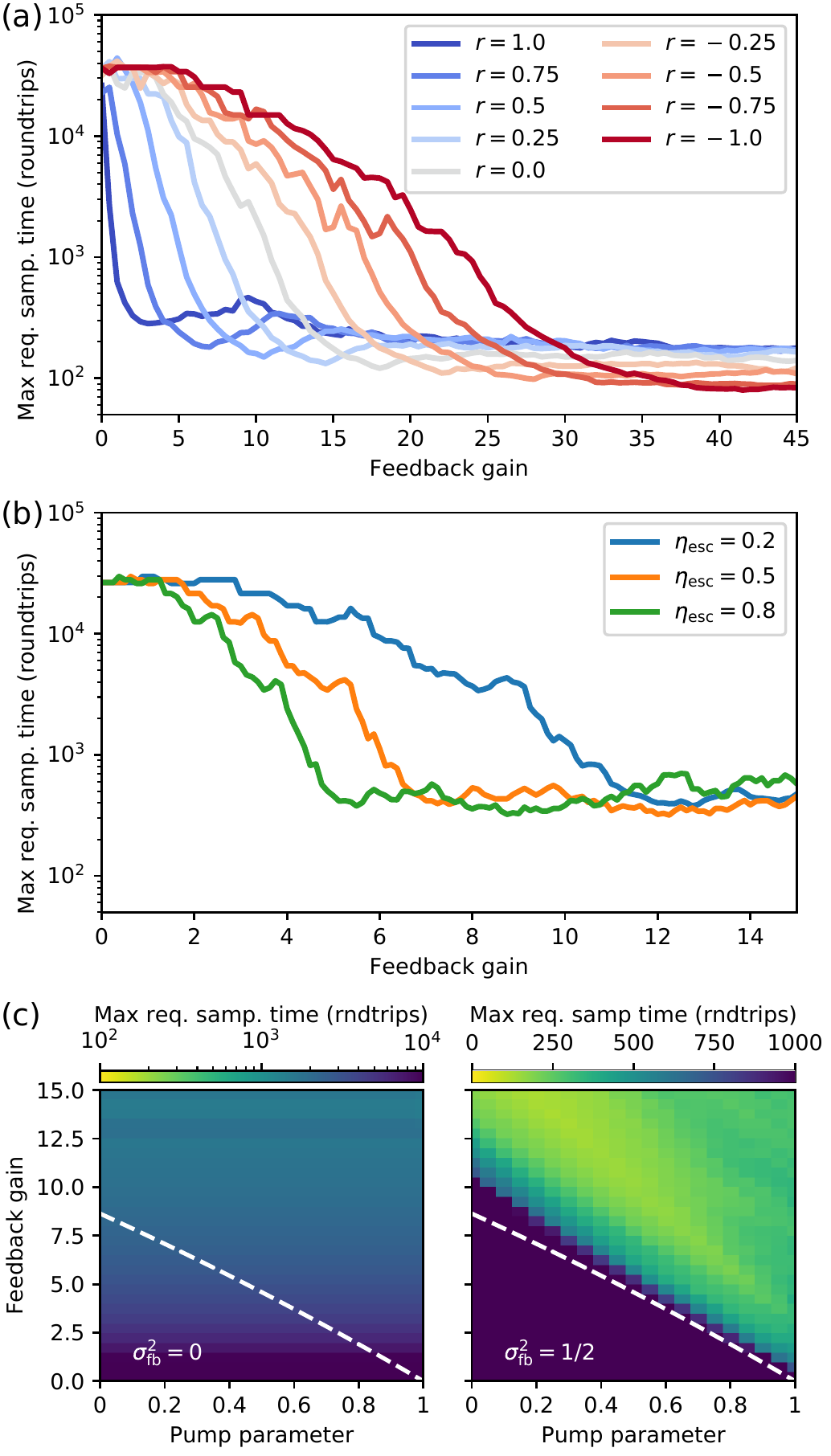} \vspace{-6pt}
    \caption{Required sampling time of various alternative models of MFB-CIM for the problem instance and spin configurations considered in Fig.~\ref{fig:sampling-metric}. (a) Required sampling time as a function of the feedback gain parameter $\alpha$ for different values of the pump parameter $r$, using the full Gaussian model of the MFB-CIM. (b) Required sampling time as a function of $\alpha$, and with various values of escape efficiency $\eta_\text{esc}$, for the MFB-CIM with no crystal ($\epsilon \tau_\text{nl} = r = 0$) and only linear dynamics. (c) Required sampling time as a function of both $r$ and $\alpha$ for the mean-field (i.e., no quantum noise) MFB-CIM, with added classical feedback noise with variance $\sigma_\text{fb}^2 = 0$ (left, i.e., no noise) and $\sigma_\text{fb}^2 = 1/2$ (right). The dashed-white line indicates the linear threshold of the system, as in Fig.~\ref{fig:sampling-param-map}. In (a) and (b), we fix $n_\text{sat} = 200$ and $\eta_\text{esc} = \num{0.2}$; the cavity decay time is set to $T_\text{decay} = \num{4}$ throughout.}
    \label{fig:alt-models}
\end{figure}

In Fig.~\ref{fig:alt-models}(a), we show the maximum required sampling time as a function of the feedback gain over a range of pump parameters $r$, including $r \leq 0$, for the previously considered case of $T_\text{decay} = \num{4}$ and $\eta_\text{esc} = \num{0.2}$ from Fig.~\ref{fig:sampling-param-map}. (For $r \geq 0$, these lines are simply vertical slices of the upper-left panel of Fig.~\ref{fig:sampling-param-map}.) We see that, surprisingly, performance is quite comparable over a wide range of pump parameters with negative $r$, corresponding to parametric \emph{deamplification}. In this regime, the system even gives slightly better performance at the expense of requiring higher feedback gain to overcome the deamplification. Generally, for sufficiently large feedback gain, there is always a robust region over which acceptable sampling performance is obtained, but for $r > 0$, there is also a ``sweet spot'' at lower feedback gain where the stochastic noise due to antisqueezing can allow for efficient sampling with lower feedback gain. In Sec.~\ref{sec:scaling} we explore how these two operational modes, $r > 0$ and $r < 0$, scale to larger problem instances.

Next, we investigate the second model where the nonlinear crystal is removed from the MFB-CIM. We set $\epsilon \tau_\text{nl} = 0$ (so $n_\text{sat} = \infty$), which eliminates the need to integrate \eqref{eq:crystal-eoms-mean} and \eqref{eq:crystal-eom-variance} for the crystal propagation each roundtrip. There is also no longer a pump parameter, leaving us with just the feedback gain parameter $\alpha$ in addition to $T_\text{decay}$ and $\eta_\text{esc}$. Note that without the nonlinear saturation, the system is unstable once the feedback gain exceeds the roundtrip attenuation due to loss and outcoupling, but because our sampling metric \eqref{eq:req-samp-time} only involves the sign of the homodyne result, the metric is unaffected so long as we terminate the simulation before numerical overflow. In Fig.~\ref{fig:alt-models}(b), we show the maximum required sampling time for this MFB-CIM model at $N_\text{decay} = 4$. We find that the performance also exhibits a certain threshold, which occurs at smaller values of feedback gain compared with the nonlinear MFB-CIM. However, the attained sampling times are greater than that of the nonlinear MFB-CIM by at least a factor of two. This observation suggests that the nonlinear saturation plays an important role in effectively embedding the Ising problem into the dynamics of the MFB-CIM, consistent with the findings of Ref.~\cite{Strinati2020}. In Sec.~\ref{sec:scaling} we explore how this model scales out to larger problem instances.

While the former two cases are straightforward to address within our model, the third approach involves taking a mean-field limit, which we can motivate as follows. For simplicity, we illustrate the limit using the continuous-time Gaussian-state EOMs \eqref{eq:high-finesse-eoms}, although by using the exact mapping detailed in Sec.~\ref{sec:cont-time-reduction}, the procedure for the discrete-time version can be similarly derived. To take the mean-field limit, we define a rescaled mean-field coordinate $\widetilde q_i \coloneqq \sqrt{g/\kappa} \vbrak{\hat q_i}$, in which case \eqref{eq:high-finesse-eoms} can be written as
\begin{subequations}\begin{align}
    \diff{\widetilde q_i}{t} &= (p-\kappa-\gamma)\widetilde q_i - \frac{\kappa}{2} \widetilde q_i{}^3 + \lambda\textstyle\sum_{j=1}^N J_{ij} \widetilde q_j \label{eq:mean-field-limit-mean}\\
    &\quad{}+ \sqrt{g}\biggl[2\paren{\vbrak{\delta\hat q_i^2} - \textstyle\frac12} \xi_i + \frac{\lambda}{\kappa} \textstyle\sum_{j=1}^N J_{ij} \xi_j\biggr], \nonumber \\
    \diff{\vbrak{\delta\hat q_i^2}}{t} &= 2p\vbrak{\delta\hat q_i^2} - 2(\gamma+\kappa)\paren{\vbrak{\delta\hat q_i^2} - \textstyle\frac12} \\
    &\quad{}- 4\kappa\paren{\vbrak{\delta\hat q_i^2} - \textstyle\frac12}^2 - 2\kappa\widetilde q_i{}^2\paren{\textstyle\frac32\vbrak{\delta \hat q_i^2}-\frac12}. \nonumber
\end{align} \end{subequations}
We now consider the limit of small single-photon nonlinearities where $g$ is very small. As long as $\widetilde q_i$ is finite, the dynamics of $\mean{\delta\hat q_i^2}$ are bounded, so the noise terms in the second line of \eqref{eq:mean-field-limit-mean} scale overall as $\sqrt{g}$, thus becoming negligible compared to the other terms of \eqref{eq:mean-field-limit-mean} in the limit $g \ll \kappa, p, \gamma, \lambda$. It also follows that $\mean{\hat q_i} \gg \mean{\delta\hat q_i^2}$, so that we can neglect the quantum fluctuations, upon which the dynamics are completely characterized by $\widetilde q_i$, i.e., the EOMs are simplified to
\begin{equation}
    \diff{\widetilde q_i}{t} = (p-\kappa-\gamma)\widetilde q_i - \frac{\kappa}{2} \widetilde q_i{}^3 + \lambda\sum_{j=1}^N J_{ij} \widetilde q_j.
\end{equation}
The numerical simulations we use to study sampling performance in this limit uses the discrete-time version of the above limit, which involve the same arguments as above. When studying mean-field dynamics, it is standard procedure to introduce a small, random initial condition to avoid unstable fixed points of the dynamics, so we also adopt this convention by setting $\widetilde q_i(0) = \sigma_i \widetilde q_0$, where we fix $\widetilde q_0 \coloneqq \num{e-3}$ and $\sigma_i$ is uniformly sampled from $\pm1$. We note the main requirement is that $\widetilde q_0$ be sufficiently small to avoid undue transients in the mean-field simulations. This can correspond, e.g., to an initial seed amplitude much smaller than those produced by the dynamics we are interested in (or indeed by any other physical effects that can destabilize an unstable fixed point).

Because $\vbrak{\delta\hat q_i^2}\!/\!\vbrak{\hat q_i}^2 \sim g/\kappa \rightarrow 0$, the fluctuations in the homodyne measurement results $\widetilde w_i \coloneqq \sqrt{g/\kappa}\, w_i$ also become negligible in this limit, and $\widetilde w_i \rightarrow r_\text{out} \widetilde q_j$. The internal cavity state, represented by simply $\widetilde q_i$, experiences no backaction (e.g., amplitude shift) upon measurement, and the feedback signal $\widetilde v_i \coloneqq J_0 \sum_{j=1}^N J_{ij} \widetilde w_j \rightarrow r_\text{out}J_0 \sum_{j=1}^N J_{ij} \widetilde q_j$ becomes a deterministic function of the internal state. If we wish to restore stochasticity while still retaining the classical character of the model, we can replace the feedback term \eqref{eq:meas-feedback} with
\begin{equation}
    \widetilde v_i \coloneqq J_0 \sum_{j=1}^N J_{ij} \paren{\widetilde w_j + z_j},
\end{equation}
where $z_j \sim \mathcal N(0,\sigma_\text{fb})$, representing the injection of classical noise into the feedback signal.

In Fig.~\ref{fig:alt-models}(c), we show the maximum required sampling time as a function of pump parameter and feedback gain for this mean-field model at $N_\text{decay} = \num{4}$ and $\eta_\text{esc} = \num{0.2}$. The left panel shows the performance of the mean-field model with $\sigma_\text{fb}^2 = 0$, as is conventionally used to study combinatorial optimization in the mean-field MFB-CIM. We find that this model is significantly less efficient at sampling than the Gaussian-state quantum model. On the other hand, setting $\sigma_\text{fb}^2 = 1/2$ in the right panel recovers much of the sampling performance of the Gaussian-state quantum model. This result suggests that efficient sampling in the MFB-CIM, while naturally accessible via quantum noise, can nevertheless be largely emulated by classical noise interacting with weak single-photon nonlinearities. Of course, this comparable performance comes at a cost: whereas $|v_i|^2$ and $|\beta|^2$ represent the approximate number of photons (i.e., quanta of energy) required to operate the feedback and pump terms of the Gaussian MFB-CIM, respectively, these energy costs are scaled by a factor of $\kappa/g$ into the large-photon-number regime for the mean-field MFB-CIM, and this is before accounting for the energy consumption, if any, associated with generating the classical noise $z_i$. Thus, despite the promising sampling performance predicted for the noisy mean-field MFB-CIM model, it incurs the cost of energy inefficiency compared to the MFB-CIM sampler driven by quantum noise.

We also remark that both the coherent-state linear model and the mean-field nonlinear model explicitly exclude, each in their own way, the quantum correlations between the internal and outcoupled pulses (i.e., $\vbrak{\hat q_i \, \hat q_\text{h}}$). Thus these two models do not feature the measurement-induced shifts in the mean and variance reduction of the internal state as described by \eqref{eq:homodyne-conditional-state} in the Gaussian model. Further research into the dynamical and operational differences among these models could help further elucidate the role of quantum effects in the mechanics of the CIM.

\begin{figure*}
    \includegraphics[width=0.98\textwidth]{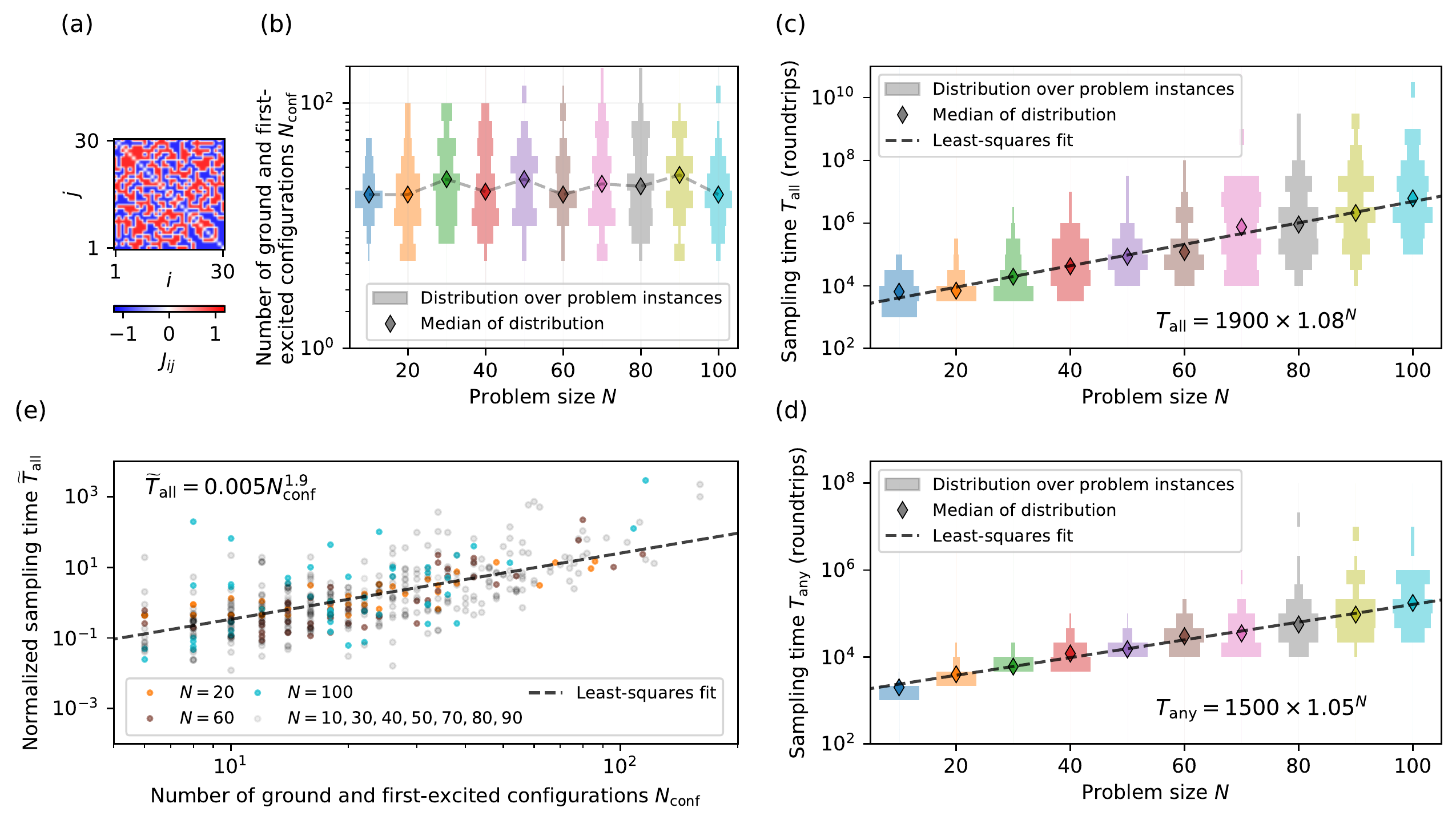}
    \vspace{-6pt}
    \caption{Scaling of the sampling performance of the MFB-CIM with respect to problem size $N$, evaluated using 50 SK1 problem instances at each $N$. (a) A representative Ising problem matrix $J_{ij}$ of the SK1 problem class at $N=30$. (b) The distribution of the number $N_\text{conf}$ of ground and first-excited configurations for the SK1 problem class for each $N$. (c) The distribution of the time $T_\text{all}$ sufficient to sample all ground and first-excited configurations for each $N$. (d) The distribution of the time $T_\text{any}$ sufficient to sample any one of the ground or first-excited configurations for each $N$. The dashed black lines represent the least-squares fit of the medians of the distributions with respect to $N$; the equations of the resultant fits are shown in the respective plots. (e) The normalized time to sample all ground and first-excited configurations as a function of $N_\text{conf}$. The normalized sampling time $\widetilde T_\text{all}$ is defined as $T_\text{all}$ divided by the median value of $T_\text{all}$, as shown in panel (c), for the respective problem size $N$. The dashed line represents a least-squares fit on all of the data with respect to $N_\text{conf}$; the equation of the resultant fit is shown in the plot. In these simulations, the MFB-CIM is operated with negative pump parameter; for more details on the model parameters used, see the ``Negative pump'' row of Table~\ref{tab:scaling}.}
    \label{fig:N-scaling}
\end{figure*}

\section{Scaling estimates of sampling performance} \label{sec:scaling}
In this section, we study the scaling of sampling performance in the discrete-time MFB-CIM with respect to problem size. We investigate the extent to which the observations and results from Sec.~\ref{sec:numerical_results} obtained from studying small and particular problem instances can generalize to larger sets of larger problems. To be concrete, we focus on the SK1 problem class introduced previously as it features instances with a large number of ground and first-excited spin configurations, and we evaluate the sampling performance of the MFB-CIM with multiple instances of this problem class at every given problem size. We also numerically study the relationship between sampling performance and the degree of degeneracy, as well as the relative scalings among the various alternative models of the MFB-CIM discussed in Sec.~\ref{sec:alternative-models}.

Here, we employ a more stringent metric than the (previously employed) required sampling time $T_\text{samp}$ to characterize the sampling performance of the MFB-CIM. Operationally, the previous metric attempts to capture a \emph{necessary} runtime for sampling, which is useful for characterizing the potential computational power of the MFB-CIM but does not prescribe a \emph{sufficient} runtime for sampling that, e.g., can be used in an experimental setting. Thus in this section, we define a sampling time $T_\text{all}$ given by the number of trajectories taken to sample all ground and first-excited configurations, multiplied by a fixed number of roundtrips $T_\text{sim}$ (i.e., the runtime) per trajectory. This definition is well suited to an experimental procedure where each trajectory is run for a predetermined, fixed time $T_\text{sim}$, so $T_\text{all}$ gives the overall time such an experiment would take. This metric is conservative in the sense that more sophisticated experimental heuristics for predicting when to stop the trajectories earlier than $T_\text{sim}$ could lead to faster sampling (bringing $T_\text{all}$ closer to $T_\text{samp}$). In addition to $T_\text{all}$, we also study the time $T_\text{any}$ to sample any one of the ground or first-excited configurations, which is similarly defined as the number of trajectories taken to sample any one of the ground or first-excited configurations, multiplied by $T_\text{sim}$.

\begin{figure*}
    \includegraphics[width=0.83\textwidth]{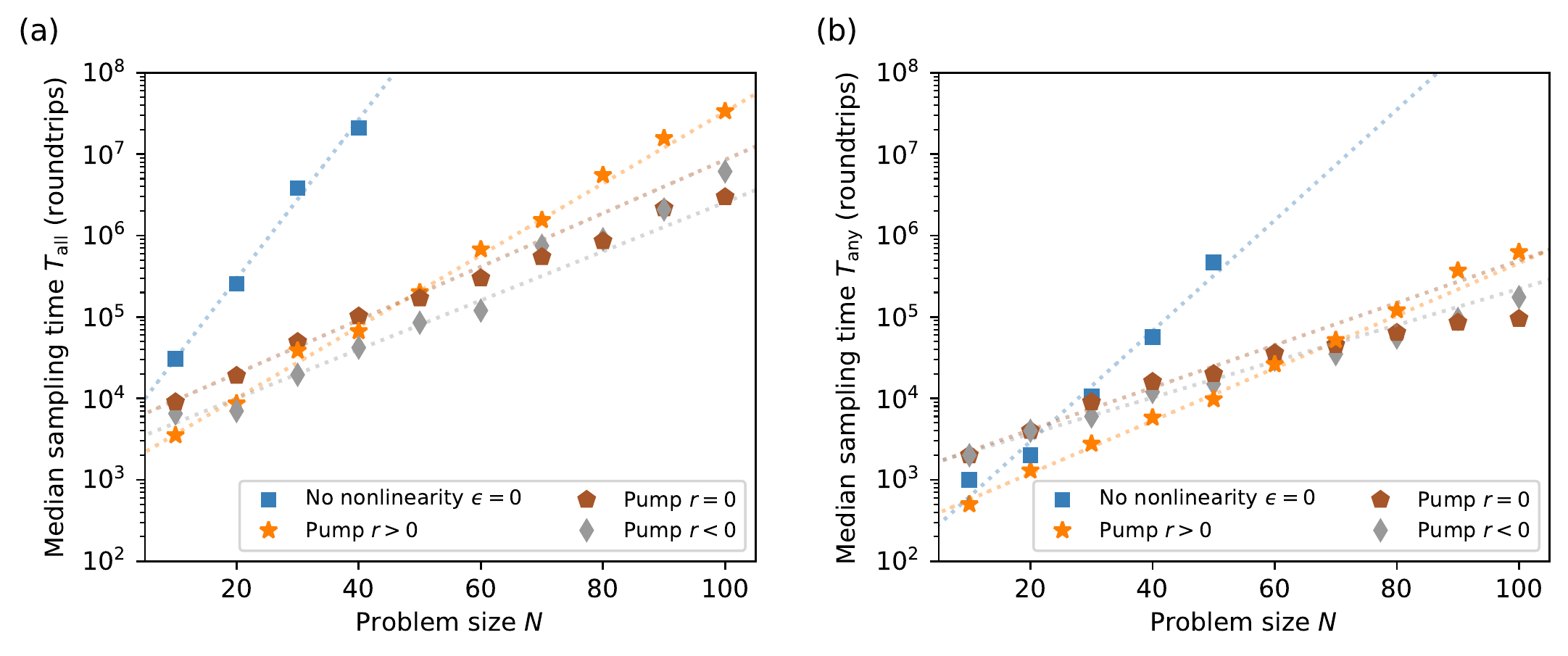}
    \caption{Sampling performance with various alternative CIM-sampling schemes, as described in Sec.~\ref{sec:alternative-models}. (a) Median time to sample all the ground and first-excited configurations as a function of problem size $N$. (b) Median time to sample any one of the ground or first-excited configurations as a function of problem size $N$. For each $N$, 50 SK1 problem instances are used considered to compute the medians, as done in Fig.~\ref{fig:N-scaling}. In both panels, the dotted lines represent least-squares fits of the median sampling times with respect to problem size. The parameters for each model are specified in Table \ref{tab:scaling}.}
    \label{fig:largeN-schemes}
\end{figure*}

Figure~\ref{fig:N-scaling} shows how the sampling performance of the MFB-CIM scales with problem size $N$. For any given $N$, we consider \num{50} problem instances from the SK1 problem class. A representative instance of this problem class can be found in Fig.~\ref{fig:N-scaling}(a), which shows that the nondiagonal elements of the problem matrix $J_{ij} \in \pm 1$. As shown in Fig.~\ref{fig:N-scaling}(b), this problem class has a large total number of degenerate ground and first-excited configurations $N_\text{conf}$, which is beneficial for evaluating sampling performance. In this paper, the degenerate ground and first-excited configurations of these problem instances have been identified using the parallel-tempering algorithm~\cite{Earl2005}. Although parallel tempering is a heuristic algorithm and does not guarantee we identified all ground and first-excited configurations, it has been shown to reliably find the ground energy for the problem instances we consider~\cite{Hamerly2016}, and, in principle, it is capable of exhaustively finding all the configurations, as the algorithm inherently produces fair samples provided it is run for a sufficiently long time (for the stochastic process to equilibrate). Figure~\ref{fig:N-scaling}(c) shows the distribution of the sampling time $T_\text{all}$ for various problem sizes. These simulations are performed with negative pump parameter ($r < 0$) since this regime was found to perform robustly for the $N=16$ instance studied in Sec.~\ref{sec:alternative-models}. Additional details about the various model parameters used are shown in Table~\ref{tab:scaling}. As stated in the table, the key parameters of $r$ and $\alpha$ are stochastically varied from trajectory to trajectory in order to account for problem-dependent variations in the dynamical threshold of the MFB-CIM, which can lead to some problem instances being stuck (an issue easy to detect and correct experimentally). Analogously to Fig.~\ref{fig:N-scaling}(c), Fig.~\ref{fig:N-scaling}(d) shows the distribution of the sampling time $T_\text{any}$, and we note the exponential scaling is consistent with prior results for the time required in Ising optimization~\cite{McMahon2016}. On the other hand, studying the scaling of $T_\text{any}$ \emph{against} that of $T_\text{all}$ provides insight into the overhead needed to sample many configurations: the difference in the base of the exponent (\num{1.05} vs \num{1.08}) suggests that while the overhead scales exponentially, the penalty (with a base $\approx \num{1.03}$) is not especially high. Finally, in Fig.~\ref{fig:N-scaling}(e), we show the scaling of the normalized sampling time $\widetilde T_\text{all}$ (specifically, normalized by the median $T_\text{all}$ at each $N$) with respect to $N_\text{conf}$. We see that there is correlation between the two quantities and that, for a fixed problem size, the sampling time scales approximately quadratically with respect to the number of configurations.

To put these results into experimental context, the sampling time in roundtrips can also be converted into wall-clock time by multiplying by the roundtrip time of the cavity, which, for a time-multiplexed MFB-CIM, is $\sim N/f_\text{rep}$. Thus, for a problem size of $N=100$ and assuming a source laser with a repetition rate of \SI{10}{GHz}, Fig.~\ref{fig:N-scaling}(c) indicates that all configurations can be sampled within a median wall-clock time of \SI{60}{\milli\second}. As a subject for future work, it would be interesting to perform more thorough benchmark studies to see how these wall-clock times compare to those attained by contemporary algorithms running on conventional digital hardware.

\setlength{\tabcolsep}{6pt}
\begin{table}
    \begin{tabular}{ l | c c c c}
         & $N_\text{decay}$ & $\alpha$ & $r$ & $\eta_\text{esc}$ \\
         \hline
         Positive pump & \num{4} & $\num{40} + \num{10}z$ & $\num{-0.8} + \num{0.08}z$ & \num{0.2} \\
         No pump & \num{4} & $\num{30} + \num{5}z$ & \num{0} & \num{0.5} \\
         Negative pump & \num{1} & $\num{4} + \num{0.6}z$ & $\num{0.8} + \num{0.05}z$ & \num{0.5} \\
         No nonlinearity & \num{2} & $\num{10} + \num{2}z$ & -- & \num{0.5} \\
    \end{tabular}
    \label{table:static-parameters}
    \caption{Model parameters used in Sec.~\ref{sec:scaling} for simulating the MFB-CIM across different operational modes (see Sec.~\ref{sec:alternative-models}) in studying the large-$N$ scaling of sampling performance. For all simulations, $T_\text{sim} = 50N_\text{decay}$, and $n_\text{sat} = 200$ for all simulations except the case of $\epsilon = 0$, where it is undefined. Here, $z \sim \mathcal N(0,1)$ denotes a standard-normal random variable that perturbs the associated parameter from one trajectory to the next according to the given formula.}
    \label{tab:scaling}
\end{table}

Lastly, we also examine how some alternative modes of operation in the MFB-CIM perform relative to each other. Figure~\ref{fig:largeN-schemes} shows, for the models considered in Sec.~\ref{sec:alternative-models}~\footnote{We do not consider the classical mean-field model here as it requires classical energy scales to realize in experiment, making the comparison unfair.}, how the medians of the time $T_\text{all}$ to sample all configurations or $T_\text{any}$ to sample any configuration scales with problem size $N$. The parameters used for each model are listed in Table~\ref{tab:scaling}; due to the large parameter space, the parameters have been heuristically chosen by optimizing over a small set (of size $\sim 10$), consisting of variations around the optimal parameters found in Sec.~\ref{sec:alternative-models} for $N = 16$. The results show that despite its decent sampling performance in the $N=16$ case, the coherent-state model scales very poorly, indicating that a linear measurement-feedback protocol in the absence of nonlinearity cannot adequately explain the sampling performance of the MFB-CIM; these results are consistent, for example, with the findings of Ref.~\cite{Strinati2020}. We also observe that setting the pump parameter to be negative or even zero results in performance that scales similarly to, or arguably even better than, the case of positive pump parameter, both for sampling all configurations as well as any. Considering the experimental advantages of no longer requiring a pump for the system, we expect the $r=0$ results to be an interesting regime to explore in MFB-CIM sampling experiments.

\section{Conclusions}

In this paper, we have formulated a numerically tractable, discrete-time model of the MFB-CIM valid down to the Gaussian-state regime in which quantum noise plays an important role in the system dynamics. Despite being based on the Gaussian-state formalism, however, the model nevertheless captures nonlinear dynamics in the mean and variance of the Gaussian state under experimentally relevant conditions by employing a second-order moment expansion to describe the propagation of the state through the intracavity nonlinearity. The resulting dynamical model is highly general, simultaneously overcoming several restrictions in previously established numerical models for MFB-CIMs: Continuous-time models based on quantum input-output theory only apply to high-finesse CIMs; linear models based on pure Gaussian operations (e.g., squeezing) only apply to CIMs operating below threshold or without optical nonlinearities; and mean-field nonlinear models only apply to high-photon-number CIMs in the classical regime where quantum noise is neglected.

The generality of our model has allowed us to examine the MFB-CIM in the context of a new computational application beyond conventional combinatorial optimization: the dynamical sampling of low-energy Ising spin configurations, driven by quantum noise. We have shown that the sampling behavior first observed in continuous-time Gaussian models of the MFB-CIM~\cite{Kako2020} persists into the low-finesse regime, carrying the important advantage of increased efficiency by bypassing the diffusive dynamics inherent to the continuous-time limit. We have provided natural parametrizations of our model of relevance to experimental settings, and we have operationally explored sampling performance across a range of these parameters, including pump rate, feedback gain, cavity finesse, and outcoupling efficiency. Using this model, we have explored different operational modes of the MFB-CIM, including negative or zero pump rates, which result in comparable or even enhanced performance, and the absence of optical nonlinearity or quantum noise, both of which result in significant degradation of sampling performance. Due to the compatibility of our model with both existing (low-finesse, high-photon-number) as well as future (quantum-noise-dominated) experimental MFB-CIMs, we expect our numerical results to have immediate implications for the path towards demonstrating efficient Ising sampling on the CIM platform.

In addition to our numerical findings in the context of Ising sampling, this paper also complements and expands upon a longstanding goal of identifying quantum mechanisms and principles of operation in the CIM~\cite{Yamamoto2020}. The ability to properly treat quantum noise in the Gaussian-state regime using a discrete-time formalism generalizes and validates previous investigations into CIM physics via continuous-time positive-P, truncated-Husimi, and truncated-Wigner SDEs~\cite{Inui2020, Inui2020b}, and it also helps clarify the limitations of mean-field models~\cite{Wang2013, Leleu2019} commonly used to study the role of nonlinear dynamics in the large-$N$ limit. Back in the small-$N$ limit, these Gaussian-regime results can act as conceptual semiclassical scaffolding on which to build better understanding of complicated and often unintuitive deep-quantum dynamics. While our focus has been on the measurement-feedback CIM in this paper for the sake of simplicity and experimental relevance, it is straightforward to generalize our approach to describe coherently-coupled CIM networks~\cite{Marandi2014b,Inagaki2016b} or potentially even other optical machines like laser networks implementing XY-spin Hamiltonians~\cite{Gershenzon2020, Pal2020}. In cases where nonlocal entanglement is generated, the cost of representing an entangled Gaussian state only scales as $\mathcal O(N^2)$, so our modeling approach can enable intermediate-$N$ numerical studies into the potential role of entanglement in these platforms.

\acknowledgments
The authors thank Yoshitaka Inui, Sam Reifenstein, Logan G.\ Wright, Ryotatsu Yanagimoto, and Evan Laksono for helpful discussions and feedback. This work was supported by the National Science Foundation under Grant No.\ CCF-1918549 and the Army Research Office under Grant No.\ W911NF-16-1-0086. P.L.M.\ acknowledges membership in the CIFAR Quantum Information Science Program as an Azrieli Global Scholar. The authors wish to thank NTT Research for their financial and technical support. E.N., T.O., S.K., and Y.Y.\ are listed as inventors on a U.S.\ provisional patent application (No.\ 63/157,673) related to this paper.

\clearpage
\bibliography{references}

\begin{appendix}

\section{Continuous-time Gaussian quantum models and the high-finesse limit}
\label{app:cont-time}

As outlined in Sec.~\ref{sec:cont-time-reduction}, the discrete-time model can be reduced to continuous-time models for CIMs derived using conventional quantum optics theory. We first give one example of such a continuous-time quantum model, which produces the Gaussian-state EOMs \eqref{eq:high-finesse-eoms} from the main text. We then analyze each of the discrete map operations described in Sec.~\ref{sec:discrete-map-recipe}, including the nonlinear crystal propagation, to show how \eqref{eq:high-finesse-eoms} can arise in the high-finesse limit of the formalism; in the process we derive the explicit relationships \eqref{eq:cont-disc-parameter-map} that characterize the scaling of all parameters in our discrete-time model required for the limit to hold. Finally, we present an alternative perspective on this limit in the language of quantum input-output theory, which may also be useful for some readers.

\subsection{Continuous-time Gaussian quantum model}

The standard approach to modeling the MFB-CIM is based on input-output theory~\cite{Wiseman2010,Gardiner1985a}, which describes open quantum systems coupled weakly to a set of external reservoirs. In this formalism, the dynamics are specified by a system Hamiltonian capturing the unitary evolution and a set of Lindblad operators, which describe the interactions of the system with the reservoirs.

For the MFB-CIM, the system of $N$ DOPOs is represented as in the discrete-time case by optical modes with annihilation operators $\hat a_i$. The system is coupled to three reservoirs. The first describes unmeasured linear loss and is represented by Lindblad operators $\hat L_{\text{loss}, i} \coloneqq \sqrt{2\gamma} \, \hat a_i$, where $\gamma$ is the field decay rate due to loss. The second describes outcoupling and is represented by Lindblad operators $\hat L_{\text{out}, i} \coloneqq \sqrt{2\kappa} \, \hat a_i$, where $\kappa$ is the field outcoupling rate. Finally, gain saturation is modeled as a two-photon loss corresponding to back-conversion of signal into pump and is represented by Lindblad operators $\hat L_{\text{tpl}, i} \coloneqq \sqrt{g} \, \hat a_i^2$, where $g$ is the two-photon loss rate.

The Hamiltonian consists of two coherent effects. The first is generated by the external pumping of the nonlinear crystal, which gives a contribution of the form $(\im p/2)\hat a_i^{\dagger2} + \text{H.c.}$, where $p$ is the field pump rate. The second is generated by external feedback injection, which is a function of the homodyne measurement record obtained from monitoring the output channels $\hat L_{\text{out}, i}$; we denote this measurement record by
\begin{align}
    m_i(t) \coloneqq \braket{\hat L_{\text{out},i} + \hat L_{\text{out},i}^\dagger} + \xi_i(t),
\end{align}
where $\xi_i(t)$ is a real-valued standard white noise process with $\delta$-function correlations $\braket{\xi_i(t) \xi_j(t')} = \delta_{ij}\delta(t-t')$. Taken together, the system Hamiltonian is given by
\begin{align}
    \label{eq:Ham-drive}
    \hat H(t) \coloneqq & \frac{\im}{2}\sum_{i=1}^N \paren{p \hat a_i^{\dagger 2} + \lambda\frac{f_i(t)}{\sqrt{2\kappa}} \hat a_i^\dagger} + \text{H.c.},
\end{align}
where $f_i(t) \coloneqq \sum_j J_{ij} m_j(t)$ is the feedback signal.

Because the measurement records $m_i(t)$ constitute continuous weak measurements of the system state, the dynamics of the system are stochastic and conditional on $m_i(t)$. In standard input-output theory, such dynamics are generated by a stochastic master equation (SME)~\cite{Wiseman1993}
\begin{align}
    \diff{\hat\rho}{t} &= -\im\Sbrak{\hat H(t), \hat\rho} + \sum_{i=1}^N \xi_i(t) \; \mathcal{H}[\hat L_{\text{out},i}] \, \hat\rho \\
    &\qquad{} + \sum_{i=1}^N \paren{\mathcal{D}[\hat L_{\text{out},i}] + \mathcal{D}[\hat L_{\text{loss},i}] + \mathcal{D}[\hat L_{\text{tpl},i}]} \hat\rho \nonumber,
\end{align}
for superoperators $\mathcal{D}[\hat A]\hat\rho \coloneqq \hat A \hat\rho {\hat A}^\dagger - \frac{1}{2}\bigl\{\hat A^\dagger\hat A,\hat\rho\bigr\}$ and $\mathcal{H}[\hat A] \hat\rho \coloneqq \bigl\{\hat A, \hat\rho\bigr\} - \bigl\langle\hat A + \hat A^\dagger\bigr\rangle\hat\rho$.

From the SME, the conditional evolution of any desired observable can be obtained. To establish a correspondence with the discrete-time model, we are particularly interested in the mean and variance of the in-phase quadrature $\hat q_i$. In general, the expectation value of an observable $\hat X$ has the equation of motion
\begin{align}
    \diff{\braket{\hat X}}{t} &= -\im\Braket{\Sbrak{\hat X, \hat H(t)}} + \sum_{i=1}^N \xi_i(t) \Braket{\mathcal{H}[\hat L_{\text{out},i}^\dagger]\hat X} \\
    &\;{} + \sum_{i=1}^N \Braket{\paren{\mathcal{D}[\hat L_{\text{out},i}^\dagger] + \mathcal{D}[\hat L_{\text{loss},i}^\dagger] + \mathcal{D}[\hat L_{\text{tpl},i}^\dagger]}\hat X} \nonumber.
\end{align}
We consider $\hat X$ to be $\hat q_i$ and $\delta\hat q_i^2$ to obtain the dynamics of the mean and variance, respectively. As in the discrete-time model, in order to arrive at a closed set of differential equations for the evolution, we assume that the state $\hat\rho$ is a Gaussian state at all times. As in the discrete-time model, this Gaussian-state approximation holds when the single-photon nonlinearity is small relative to the linear loss/measurement rates, i.e., $g \ll \kappa + \gamma$. As shown in Appendix~\ref{app:weyl}, expectation values of quadrature operators can be evaluated under the Gaussian-state assumption. Using the procedure outlined there, we arrive at
\begin{subequations} \label{eq:exact-high-finesse-eoms}
\begin{align}
    \diff{\braket{\hat{q}_i}}{t} &= (p-\kappa-\gamma)\braket{\hat q_i} - \frac{g}{2}\braket{\hat q_i}^3 + \lambda \sum_{j=1}^N J_{ij} \braket{\hat q_j} \nonumber\\
    &{} + 2\sqrt{\kappa}\paren{\braket{\delta\hat q^2_i} - \textstyle\frac{1}{2}} \xi_i + \frac{\lambda}{2\sqrt{\kappa}} \sum_{j=1}^N J_{ij} \xi_j \\
    &{} - \frac{g}{2}\braket{\hat q_i}\paren{3\braket{\delta \hat q_i^2} + \braket{\delta \hat p_i^2} - 2}, \nonumber\\
    \diff{\braket{\delta\hat{q}^2_i}}{t} &= 2(p-\kappa-\gamma)\braket{\delta\hat q^2_i} - 4\kappa\paren{\braket{\delta\hat q^2_i} - \textstyle\frac{1}{2}}^2 \nonumber\\
    &{} + \kappa + \gamma - 3 g\braket{\hat q_i}^2\braket{\delta\hat q^2_i} + g\braket{\hat q_i}^2 \\
    &{} - g\sbrak{3\braket{\delta\hat q^2_i}\paren{\braket{\delta\hat q^2_i} +                  \textstyle\frac13\braket{\delta\hat p^2_i} - 1} - \braket{\delta\hat p^2_i} + 1}\nonumber.
\end{align}
\end{subequations}
For the Gaussian-state approximation to be valid, we require $g \ll \kappa + \gamma$. Thus, terms that scale as $g$ should only be kept if the factor accompanying the $g$ has the capacity to be large. This is for instance satisfied in the saturation term $-g \braket{\hat q_i}^3$, where a large displacement $\braket{\hat q_i}$ can make it comparable to the other terms such as $(p-\kappa-\gamma)\braket{q_i}$. Accordingly, we see that the final terms of both equations in \eqref{eq:exact-high-finesse-eoms} can in fact be neglected as having loss and measurement $\kappa + \gamma \gg g$ ensures the amount of squeezing/antisqueezing in the MFB-CIM is modest. Removing those terms, we arrive at the simplified continuous-time Gaussian model \eqref{eq:high-finesse-eoms}.

\subsection{High-finesse limit of discrete-time dynamics}

We now show that continuous-time dynamics of the form \eqref{eq:high-finesse-eoms} can be obtained from the discrete-time model in the high-finesse limit, where each discrete operation in the MFB-CIM only effects a small change to the state.

As discussed in Sec.~\ref{sec:cont-time-reduction} of the main text, the high-finesse limit can be defined by the limit $\delta\rightarrow 0$, where $\delta$ scales the parameters of our discrete-time model according to \eqref{eq:cont-map-assump}. We now consider each of the operations in Sec.~\ref{sec:discrete-map-recipe} and expand each of them up to first order in $\delta$. As usual, the $q$- and $p$-quadratures of the dynamics are decoupled, so we only consider the dynamics of $\hat q_i$ below.

First, we consider the linear loss at the facets given by \eqref{eq:facet-loss-map}. Using \eqref{eq:beamsplitter-map}, this produces the mapping
\begin{subequations}
\begin{align}
    \vbrak{\hat q_i} &\mapsto \sqrt{1-r_\text{loss}^2}\vbrak{\hat q_i} \nonumber\\
    &= \paren{1 - \textstyle\frac 1 2 r_\text{loss}^2}\vbrak{\hat q_i} + \mathcal O(\delta^2), \\
    \vbrak{\delta\hat q_i^2} &\mapsto  \paren{1-r_\text{loss}^2}\vbrak{\delta\hat q_i^2} + \textstyle\frac 1 2 r_\text{loss}^2.
\end{align}
\end{subequations}
Since there are two of these facets in a given round-trip, cascading the discrete map twice gives 
\begin{subequations} \label{eq:high-finesse-loss}
\begin{align}
    \vbrak{\hat q_i} &\mapsto \paren{1 - r_\text{loss}^2}\vbrak{\hat q_i} + \mathcal O(\delta^2), \\
    \vbrak{\delta\hat q_i^2} &\mapsto  \paren{1-2r_\text{loss}^2}\vbrak{\delta\hat q_i^2} + r_\text{loss}^2 + \mathcal O(\delta^2).
\end{align}
\end{subequations}

Second, we consider the crystal propagation. Since the map \eqref{eq:crystal-map-abstract} requires integrating the nonlinear EOMs \eqref{eq:crystal-eoms-mean} and \eqref{eq:crystal-eom-variance}, we use Picard iteration to solve the EOMs while only keeping terms at $\mathcal O(\delta)$; the result is an analytic map for $\vbrak{\hat q_i}$ and $\vbrak{\delta\hat q_i^2}$ correct up to $\mathcal O(\delta)$. With Picard iteration starting from the initial conditions \eqref{eq:ode-init-conditions}, we find that the crystal propagation in the high-finesse limit produces
\begin{subequations} \label{eq:high-finesse-crystal}
\begin{align}
    \vbrak{\hat q_i} &\mapsto\textstyle \vbrak{\hat q_i} + \frac1{\sqrt2}\beta \epsilon \tau_\text{nl} \vbrak{q_i} - \frac18(\epsilon \tau_\text{nl})^2 \vbrak{\hat q_i}^3 \\
    &\textstyle\quad{}- \frac18(\epsilon \tau_\text{nl})^2 \vbrak{\hat q_i}\paren{3\vbrak{\delta q_i^2} + \vbrak{\delta\hat p_i^2} - 2} + \mathcal O(\delta^2), \nonumber \\
    \vbrak{\delta\hat q_i^2} &\mapsto\textstyle \vbrak{\delta\hat q_i^2} + \sqrt2\beta\epsilon \tau_\text{nl} \vbrak{\delta\hat q_i^2} \\
    &\textstyle\quad{} - \frac34(\epsilon \tau_\text{nl})^2 \vbrak{\hat q_i}^2\vbrak{\delta\hat q_i^2} + \frac 1 4(\epsilon \tau_\text{nl})^2 \vbrak{\hat q_i}^2 \nonumber\\
    &\textstyle\quad{} - \frac14(\epsilon \tau_\text{nl})^2 \vbrak{\delta q_i^2}\paren{\vbrak{\delta\hat q_i^2} - \vbrak{\delta p_i^2}} + \mathcal O(\delta^2) \nonumber.
\end{align}
\end{subequations}
We see the last terms in both of the above equations scale as $(\epsilon\tau_\text{nl})^2$ and only occur with low powers of the mean $\braket{\hat q_i}$. We can therefore neglect them following the same argument used above for eliminating the last terms of \eqref{eq:exact-high-finesse-eoms}: Since the Gaussian-state approximation requires $(\epsilon\tau_\text{nl})^2 \ll r_\text{loss}^2 + r_\text{out}^2$, the outcoupling and loss keep the variances close to unity, thus ensuring that these terms remain much smaller than terms at the same order in $\mean{\hat q_i}$ but associated with $r_\text{loss}^2$ and $r_\text{out}^2$.

Third, we consider the measurement process. This consists first of an outcoupling step, which changes the signal state according to
\begin{subequations} \label{eq:high-finesse-outcoupling}
\begin{align}
    \vbrak{\hat q_i} &\mapsto \paren{1 - \textstyle\frac 1 2 r_\text{out}^2}\vbrak{\hat q_i} + \mathcal O(\delta^2), \\
    \vbrak{\delta\hat q_i^2} &\mapsto  \paren{1-r_\text{out}^2}\vbrak{\delta\hat q_i^2} + \textstyle\frac 1 2 r_\text{out}^2,
\end{align}
\end{subequations}
and also produces a weak correlation between $\hat q_i$ and an external mode (labeled here by a subscript $h$), with mean, variance, and covariance,
\begin{subequations}
\begin{align}
    \vbrak{\hat q_\text{h}} &= r_\text{out} \vbrak{\hat q_i}, \\
    \vbrak{\delta\hat q_\text{h}^2} &= \textstyle\frac 1 2 + r_\text{out}^2\paren{\vbrak{\delta\hat q_i^2} - \frac 1 2}, \\
    \vbrak{\delta\hat q_i\delta\hat q_\text{h}} &= \textstyle r_\text{out}\sqrt{1-r_\text{out}^2} \paren{\vbrak{\delta\hat q_i^2} - \frac 1 2}.
\end{align}
\end{subequations}
After this, the outcoupled field is measured by homodyne, which by \eqref{eq:homodyne-record} produces a measurement result
\begin{align} \label{eq:homodyne-result}
    w_i &= \mathcal N\Paren{\vbrak{\hat q_\text{h}}, \vbrak{\delta\hat q_\text{h}^2}} \\
    &= r_\text{out}\mean{\hat q_i} + \textstyle\frac1{\sqrt2}\Sbrak{1 + r_\text{out}^2\paren{\mean{\delta\hat q_i^2}-\frac12}}z_i + \mathcal O(\delta^2),\nonumber
\end{align}
where $z_i \sim \mathcal N(0,1)$ is a standard normal random variable. At the same time, backaction on the internal state by \eqref{eq:homodyne-conditional-state} produces the map
\begin{subequations} \label{eq:high-finesse-backaction}
\begin{align}
    \vbrak{\hat q_i} &\mapsto \vbrak{\hat q_i} + \paren{\frac{w_i-\vbrak{\hat q_\text{h}}}{\vbrak{\delta\hat q_\text{h}^2}}} \vbrak{\delta\hat q_i\delta\hat q_\text{h}} \nonumber\\
    &= \vbrak{\hat q_i} + \sqrt2r_\text{out}\paren{\textstyle\vbrak{\delta\hat q_i^2} - \frac 1 2} z_i + \mathcal O(\delta^{3/2}), \\
    \vbrak{\delta\hat q_i^2} &\mapsto \vbrak{\delta\hat q_i^2} - \frac{\vbrak{\delta\hat q_i\delta\hat q_h}^2}{\vbrak{\delta\hat q_h^2}} \nonumber\\
    &= \vbrak{\delta\hat q_i^2} - 2r_\text{out}^2 \paren{\textstyle\vbrak{\delta\hat q_i^2} - \frac 1 2}^2 + \mathcal O(\delta^2).
\end{align}
\end{subequations}

Finally, we consider injection feedback via \eqref{eq:meas-feedback}. Given the measurement results \eqref{eq:homodyne-result}, the displacement we apply is given by $v_i = J_0 \sum_{j=1}^N J_{ij} w_j$, which produces
\begin{align} \label{eq:high-finesse-feedback}
    \vbrak{\hat q_i} &\mapsto \vbrak{\hat q_i} + J_0 \sum_{j=1}^N J_{ij} w_j \\
    &= \vbrak{\hat q_i} + J_0 \sum_{j=1}^N J_{ij} \paren{r_\text{out}\vbrak{\hat q_i} + \textstyle\frac{1}{\sqrt 2}z_i} + \mathcal O(\delta^{3/2}),\nonumber
\end{align}
and the variance $\vbrak{\delta\hat q_i^2}$ is unchanged by the feedback.

We can now finally put together all the maps within a single roundtrip by summing up the contributions of \eqref{eq:high-finesse-loss}, \eqref{eq:high-finesse-crystal}, \eqref{eq:high-finesse-outcoupling}, \eqref{eq:high-finesse-backaction}, and \eqref{eq:high-finesse-feedback} up to first order in $\delta$. The resulting updated state described by $\vbrak{\hat q_i}'$ and $\Vbrak{\delta\hat q_i^2}'$ can then be substituted into the definitions \eqref{eq:cont-disc-relation}. Then by imposing conditions related to the Gaussian-state approximation (due to $(\epsilon \tau_\text{nl})^2 \ll r_\text{loss}^2 + r_\text{out}^2$) as was also done in the derivation of \eqref{eq:high-finesse-eoms}, we finally arrive at EOMs identical to \eqref{eq:high-finesse-eoms}, \emph{provided} we utilize the relationships explicitly given in \eqref{eq:cont-disc-parameter-map}.

\subsection{Quantum input-output approach}

Finally, as an alternative to the above approach where the correspondence between continuous and discrete time is made via manipulation of the (c-number) means and variances, it is also possible to arrive at the same conclusions using a quantum input-output analysis of the crystal Hamiltonian \eqref{eq:crystal-hamiltonian} as well. For example, on one roundtrip, the crystal implements a unitary operation
\begin{align}
    \hat U_\text{nl} \coloneqq \e{-\im \hat H_\text{nl}\tau_\text{nl}} = \exp\paren{\frac{\epsilon \tau_\text{nl}}{2}\sum_i  \hat b_i \hat a_i^{\dagger 2}  - \text{H.c.}},
\end{align}
where $\hat H_\text{nl} \coloneqq \sum_i \hat H^{(i)}_\text{nl}$. We can decompose the pump operator as the sum of a coherent-excitation part and a quantum noise part via $\hat b_i = \beta_i/\sqrt{2} + \delta\hat b_i$, allowing us to treat the parametric amplification and the nonlinear parametric quantum fluctuations separately. With this substitution,
\begin{small}\begin{align}
    \hat U_\text{nl} = \exp\paren{\frac{\beta \epsilon \tau_\text{nl}}{2\sqrt{2}}\sum_i  \hat a_i^{\dagger 2}  + \frac{\epsilon \tau_\text{nl}}{2}\sum_i \delta\hat b_i  \hat a_i^{\dagger 2} - \text{H.c.}}.
\end{align}\end{small}

In the high-finesse limit where $\beta^2 \sim (\epsilon \tau_\text{nl})^2 \sim \Delta t \sim \delta$, this unitary evolution can be made compatible with a discrete-time picture of the dynamics if we Trotterize~\cite{Sakurai2017} the above unitary over one roundtrip time by writing
\begin{subequations}
\begin{small}\begin{equation}
    \hat U_\text{nl} = \exp\paren{-\im \hat H_\text{sqz} \Delta t} \exp\paren{-\im \hat H_\text{tpl} \Delta t} + \mathcal O(\delta^{3/2}),
\end{equation}\end{small}
where the first exponential effects a rotation $\mathcal O(\delta)$ and is generated by a squeezing Hamiltonian
\begin{align} \label{eq:squeezing-hamiltonian-correspondence}
    \hat H_\text{sqz} \coloneqq \frac{\im}{2}\underbrace{\frac{\beta \epsilon \tau_\text{nl}}{\sqrt{2}\Delta t}}_{p}\sum_i  \hat a_i^{\dagger 2} + \text{H.c.},
\end{align}
while the second exponential effects a rotation $\mathcal O(\delta^{1/2})$ and is generated by an interaction Hamiltonian that we can write as
\begin{align} \label{eq:tpl-hamiltonian-correspondence}
    \hat H_\text{tpl} \coloneqq \im\underbrace{\frac{\epsilon \tau_\text{nl}}{2 \sqrt{\Delta t}}}_{\sqrt{g}}\sum_i \hat{b}^{(\text{in},t)}_i \hat a_i^{\dagger 2} + \text{H.c.},
\end{align}
\end{subequations}
where in the limit $\Delta t \sim \delta \rightarrow 0$, the quantum white-noise operators $\hat{b}^{(\text{in},t)}_i \coloneqq \delta\hat b_i/ \sqrt{\Delta t}$ have Dirac-delta commutation relations, i.e., $\Sbrak{\hat{b}^{(\text{in},t)}_i, \hat{b}^{(\text{in},t')\dagger}_{i'}} = \delta_{i,i'}\delta(t-t')$.

In a coarse-grained continuous-time theory over many roundtrips, \eqref{eq:squeezing-hamiltonian-correspondence} is precisely the gain/squeezing part of the continuous-time system Hamiltonian (23), while \eqref{eq:tpl-hamiltonian-correspondence} is an input-output system-reservoir interaction Hamiltonian that formally defines the continuous-time Lindblad operator $\hat L_{\text{tpl},i}$ in the continuous-time model. This process of Trotterizing discrete-time operations can also be applied to all the linear operations (loss, outcoupling, measurement, and feedback) as well.

\section{Evaluating expectation values of quadrature operators}
\label{app:weyl}
Here we outline how to evaluate expectation values of an operator of the form $\hat q^r \hat p^m$ on a single-mode Gaussian state. We include this section largely for pedagogical purposes as we found these results are typically presented in more general, and hence less accessible, terms than necessary for the specific scenario we consider~\cite{Schork2015}.

First we note the Weyl-ordered (i.e., symmetrically ordered) expression for the operator is given by
\begin{align}
    \hat q^r \hat p^m = \sum_{j=0}^{\text{min}(r, m)} \paren{\frac{\im}{2}}^2\binom{r}{j}\binom{m}{j} j!\paren{\hat q^{r-j} \hat p^{m-j}}_\text{W},
\end{align}
where $(\cdot)_\text{W}$ denotes operators that are Weyl-ordered~\cite{Schork2015}. With the operators in Weyl form, the expectation value can be evaluated by a phase-space integral
\begin{align}
    \vbrak{\paren{\hat q^{r-j} \hat p^{m-j}}_\text{W}} &= \int_{\mathbb{R}^2} q^{r-j} p^{m-j} W(q, p) \,\dif q \,\dif p,
\end{align}
where $W(q,p)$ is the Wigner function for the quantum state. For a Gaussian state, the Wigner function is given by a multivariate normal distribution:
\begin{align}
    W(z) = \frac{1}{2\pi\det\Sigma} \exp\paren{-\frac 1 2 (z-\mu)^\mathrm{T} \Sigma^{-1} (z-\mu)},
\end{align}
where $z = (q,p)^\mathrm{T}$ and $\mu$ and $\Sigma$ are the mean vector and covariance matrix of the state, as defined in \eqref{eq:mean}.
\vspace{14pt}

\section{Crystal propagation EOMs}
\label{app:full-eoms}

Here we present the full equations of motion for the evolution of the Gaussian moments for the joint signal-pump state as it propagates through the crystal according to \eqref{eq:crystal-hamiltonian}, using the procedure described in Sec.~\ref{sec:crystal-propagation}.

The full mean-field equations of motion are
\begin{align*}
    \diff{\mean{\hat x_i}}{(\epsilon\tau)} &= \mean{\hat x\pump}\mean{\hat x_i} + \mean{\hat y\pump}\mean{\hat y_i} + \mean{\delta\hat x\pump\,\delta\hat x_i + \delta\hat y\pump\,\delta\hat y_i}, \\
    \diff{\mean{\hat x\pump}}{(\epsilon\tau)} &= -\frac12 \Paren{\mean{\hat x_i}^2 - \mean{\hat y_i}^2} - \frac12 \mean{\delta\hat x_i^2 - \delta\hat y_i^2}, \\
    \diff{\mean{\hat y_i}}{(\epsilon\tau)} &=\mean{\hat y\pump}\mean{\hat x_i} - \mean{\hat x\pump}\mean{\hat y_i} + \mean{\delta\hat y\pump\,\delta\hat x_i - \delta\hat x\pump\,\delta\hat y_i}, \\
    \diff{\mean{\hat y\pump}}{(\epsilon\tau)} &= -\mean{\hat x_i}\mean{\hat y_i} - \frac12 \mean{\delta\hat x_i\,\delta\hat y_i + \delta\hat y_i\,\delta\hat x_i},
\end{align*}
while for the covariances, we have
\begin{widetext}
\begin{align*}
    \diff{\mean{\delta\hat x_i^2}}{(\epsilon\tau)} &=+2\mean{\hat x\pump}\mean{\delta\hat x_i^2} + 2\mean{\hat x_i}\covar{\hat x\pump}{\hat x_i} + 2\mean{\hat y_i}\covar{\hat y\pump}{\hat x_i} + \mean{\hat y\pump}\mean{\delta\hat x_i\,\delta\hat y_i + \delta\hat y_i\,\delta x_i}, \\
    \diff{\mean{\delta\hat y_i^2}}{(\epsilon\tau)} &= -2\mean{\hat x\pump}\mean{\delta\hat y_i^2} + 2\mean{\hat x_i}\covar{\hat y\pump}{\hat y_i} - 2\mean{\hat y_i}\covar{\hat x\pump}{\hat y_i} + \mean{\hat y\pump}\mean{\delta\hat x_i\,\delta\hat y_i + \delta\hat y_i\,\delta x_i}, \\
    \diff{\mean{\delta\hat x\pump^2}}{(\epsilon\tau)} &= -2\mean{\hat x_i}\covar{\hat x\pump}{\hat x_i} - 2\mean{\hat y_i}\covar{\hat x\pump}{\hat y_i}, \\
    \diff{\mean{\delta\hat y\pump^2}}{(\epsilon\tau)} &= -2\mean{\hat x_i}\covar{\hat y\pump}{\hat y_i} - 2\mean{\hat y_i}\covar{\hat y\pump}{\hat x_i}, \\
    \diff{\covar{\hat x\pump}{\hat x_i}}{(\epsilon\tau)} &= +\mean{\hat x\pump}\covar{\hat x\pump}{\hat x_i} + \mean{\hat x_i}\mean{\delta x\pump^2 - \delta x_i^2} + \mean{\hat y\pump}\covar{\hat x\pump}{\hat y_i} + \mean{\hat y_i}\mean{\delta\hat x\pump\,\delta\hat y\pump + \delta y_i\,\delta x_i}, \\
    \diff{\covar{\hat y\pump}{\hat y_i}}{(\epsilon\tau)} &= -\mean{\hat x\pump}\covar{\hat y\pump}{\hat y_i} + \mean{\hat x_i}\mean{\delta\hat y\pump^2 - \delta y_i^2} + \mean{\hat y\pump}\covar{\hat y\pump}{\hat x_i} - \mean{\hat y_i}\mean{\delta\hat x_i\,\delta\hat y_i + \delta\hat y\pump\delta\hat x\pump}, \\
    \diff{\covar{\hat x_i}{\hat y_i}}{(\epsilon\tau)} &= +\mean{\hat x_i}\mean{\delta\hat x\pump\,\delta\hat y_i + \delta\hat y\pump\,\delta\hat x_i} - \mean{\hat y_i}\mean{\delta\hat x\pump\,\delta\hat x_i - \delta\hat y\pump\,\delta\hat y_i} + \mean{\hat y\pump}\mean{\delta\hat x_i^2 + \delta\hat y_i^2},
\end{align*}
\begin{align*}
    \diff{\covar{\hat x\pump}{\hat y\pump}}{(\epsilon\tau)} &= -\mean{\hat x_i}\mean{\delta\hat x\pump\,\delta\hat y_i + \delta\hat y\pump\,\delta\hat x_i} - \mean{\hat y_i}\mean{\delta\hat x\pump\,\delta\hat x_i - \delta\hat y\pump\,\delta\hat y_i}, \\
    \diff{\covar{\hat x\pump}{\hat y_i}}{(\epsilon\tau)} &= -\mean{\hat x\pump}\covar{\hat x\pump}{\hat y_i} + \mean{\hat x_i}\mean{\delta\hat x\pump\,\delta\hat y\pump - \delta\hat x_i\,\delta\hat y_i} + \mean{\hat y\pump}\covar{\hat x\pump}{\hat x_i} - \mean{\hat y_i}\mean{\delta\hat x\pump^2 - \delta\hat y_i^2}, \\
    \diff{\covar{\hat y\pump}{\hat x_i}}{(\epsilon\tau)} &= +\mean{\hat x\pump}\covar{\hat y\pump}{\hat x_i} + \mean{\hat x_i}\mean{\delta\hat x\pump\,\delta\hat y\pump - \delta\hat x_i\,\delta\hat y_i} + \mean{\hat y\pump}\covar{\hat y\pump}{\hat y_i} + \mean{\hat y_i}\mean{\delta\hat y\pump^2 - \delta\hat x_i^2}.
\end{align*}
\end{widetext}

\end{appendix}

\end{document}